\documentclass[12pt,a4paper]{article}
 
\usepackage{amsmath,amssymb,amsthm}
\usepackage{graphicx}
\usepackage{hyperref}
\usepackage{cite}
\usepackage{physics}       
\usepackage{dcolumn}
\usepackage{braket}
\usepackage[margin=2.5cm]{geometry}
\usepackage{xcolor}
\usepackage{float}
 
\newtheorem{proposition}{Proposition}

\newcommand{\IOH}{\mathrm{IOH}}
\newcommand{\KG}{\mathrm{KG}}
\newcommand{\half}{\tfrac{1}{2}}

\begin{document}
 
\title{Quantum and Thermal Properties of the Klein-Gordon
       Inverted Harmonic Oscillator with Physical Applications}
 
\author{K. Hernández$^1$ \and M. Maamache$^2$}
 
\date{%
  $^{1}$Escuela de Física, Facultad de Ciencias Naturales y Matemáticas, Universidad de El Salvador, Final de Av. Mártires y Héroes del 30 de julio, San Salvador, El Salvador\\
  \texttt{kevinhernandezbel@hotmail.com}\\
    $^{2}$Laboratoire de Physique Quantique et Syst`emes Dynamiques,
Facult´e des Sciences, Ferhat Abbas S´etif 1, S´etif 19000, Algeria\\
  \texttt{maamache@univsetif.dz}\\
  \today
}
 
\maketitle
  
\begin{abstract}
We develop a systematic framework for the quantum and thermal
properties of a Klein-Gordon scalar field subject to an inverted
harmonic potential $-\half m^2\omega^2 x^2$.
Starting from a non-Hermitian momentum substitution
$P \to P - m\omega x$, we employ a symplectic phase-space rotation
$V = \exp\!\left[-\tfrac{\pi}{8}(xp+px)\right]$ to map the system
onto an analytically tractable effective harmonic oscillator
evaluated at $xe^{i\pi/4}$.
This allows us to define a well-regulated partition function
$Z(\beta,\omega,m)$ and derive closed-form expressions for the
free energy, entropy, and thermal correlation functions.
We then apply this framework to three physical settings:
(i) scalar field fluctuations during cosmological inflation,
(ii) quantum fields near black-hole horizons, and
(iii) order-parameter dynamics near second-order phase transitions
in condensed matter.
Our results unify previously scattered results in the literature
and provide new predictions for the finite-temperature spectral
density and entanglement entropy of unstable quantum systems.
\end{abstract}
 
 
 

\section{Introduction}
\label{sec:intro}
 
The inverted harmonic oscillator (IOH), defined by the potential
$V(x) = -\frac{1}{2}m\omega^2 x^2$, occupies a singular position
in quantum mechanics: it is one of the few exactly solvable systems
whose classical dynamics is intrinsically unstable, yet whose quantum
structure is remarkably rich and far-reaching.
Since the foundational work of Barton~\cite{Barton1986}, who established
that the spectrum is continuous and the eigenfunctions are expressed in
terms of parabolic cylinder functions, the IOH has served as a theoretical
laboratory connecting quantum mechanics, field theory, and
gravity~\cite{subramanyan2021physics}.
 
Unlike the standard harmonic oscillator, the IOH does not admit a
discrete, bounded-below spectrum, which immediately raises non-trivial
questions about the definition of quantum states, the role of
boundary conditions in the complex plane, and the physical interpretation
of its eigenfunctions~\cite{finster2017lp, liu2025quantum}.
These subtleties have attracted sustained attention:
the connection between the IOH and the Riemann zeta
function~\cite{bhaduri1997riemann}, its duality with inverse-square
potentials~\cite{sundaram2024duality}, and its role as a paradigm for
quantum chaos and complexity~\cite{bhattacharyya2021multi,
hashimoto2020exponential} all point to a system whose apparent
simplicity conceals deep mathematical and physical structure.
 
On the physical side, the IOH appears naturally wherever a quantum
field sits near the top of a potential barrier or at an unstable
equilibrium point.
In cosmology, the inflaton field near the top of its
potential~\cite{Starobinsky1980, Guth1981, Linde1982} is locally
described by an inverted harmonic potential, making the IOH directly
relevant to the physics of inflation and primordial particle
production~\cite{ParkerParticle1969, boisseau2015scalar}.
Near the event horizon of a black hole, the effective radial potential
experienced by a scalar field acquires precisely the inverted-oscillator
form~\cite{Hawking1975, Unruh1976}, linking the IOH to Hawking radiation
and the information paradox.
In condensed matter, the soft mode of a second-order phase transition
becomes massless at the critical point and is governed, in the ordered
phase, by an inverted potential~\cite{Sachdev2011, gietka2021inverted},
while more recently the IOH has been identified as the effective
Hamiltonian governing the nonequilibrium dynamics of the Dicke
model~\cite{gietka2021inverted}.
 
The study of thermal and quantum-statistical properties of the IOH
has gained considerable momentum in recent years.
Thermal properties of the IOH confined in a quantum
well~\cite{wang2025thermodynamics} and in the presence of magnetic
fields with applications to polyatomic
molecules~\cite{ngiangia2025thermal} have been investigated,
as have the adiabatic and instantaneous transitions between the
harmonic and inverted-harmonic
regimes~\cite{dodonov2024adiabatic}.
Nevertheless, a systematic \emph{field-theoretic} treatment of the
thermal and quantum properties of a scalar field subject to an
inverted harmonic potential --- grounded in the Klein-Gordon equation
and valid across the physical applications mentioned above --- is
still lacking in the literature.
 
The central difficulty is well known: a naïve evaluation of the
partition function $Z = \mathrm{Tr}(e^{-\beta H})$ diverges because the
IOH spectrum is unbounded below.
Regularisation schemes exist but are often introduced on a
case-by-case basis without a unifying justification.
A key observation, explored within the framework of
$\mathcal{PT}$-symmetric quantum
mechanics~\cite{Bender1998, Bender2007, Mostafazadeh2002},
is that the substitution $P \to P - m\omega x$, combined with a
symplectic phase-space rotation belonging to $Sp(2,\mathbb{R})$,
maps the IOH onto an effective harmonic oscillator evaluated at a
complex argument.
This mapping yields a discrete effective spectrum, renders the
partition function well defined without ad-hoc cutoffs, and
provides a natural bridge between the non-Hermitian structure of
the IOH and the standard apparatus of thermal field
theory~\cite{Kapusta2006, LeBellac1996, Matsubara1955}.
 
In this work we develop this program systematically.
Starting from the Klein-Gordon equation with an inverted harmonic
potential, we introduce the dilation operator
\begin{equation}
  V = \exp\!\left[-\frac{\pi}{8}(xp + px)\right],
  \label{eq:V_intro}
\end{equation}
whose action on the canonical pair $(x,p)$ is a rotation by
$\pi/4$ in the complex phase space~\cite{Yuen1976, Arvind1995}.
Under $V$, the Klein-Gordon inverted oscillator (KG-IOH) reduces to
an effective harmonic oscillator with known, analytically exact
solutions expressed in terms of Hermite polynomials evaluated at
$xe^{i\pi/4}$~\cite{Abramowitz, DLMF}.
This effective oscillator carries a discrete spectrum from which we
construct the partition function, the free energy, the entropy,
and the finite-temperature two-point functions of the theory.
 
The framework developed here is then applied to three distinct
physical settings:
\begin{enumerate}
  \item \textbf{Cosmological inflation.}
        The inflaton near the top of its potential is modelled as
        a KG-IOH scalar field. We compute the effective temperature
        and the thermal power spectrum of fluctuations.
 
  \item \textbf{Black-hole horizons.}
        The near-horizon geometry produces an effective inverted
        potential for a minimally coupled scalar field. Our thermal
        formalism recovers the Hawking temperature as a special case
        and provides corrections from the full KG-IOH structure.
 
  \item \textbf{Second-order phase transitions.}
        The critical soft mode is identified with the KG-IOH field.
        We derive the equation of state and the thermal scaling of
        the order parameter near the critical point.
\end{enumerate}
 
The paper is organised as follows.
Section~\ref{sec:qm} reviews the quantum mechanics of the IOH,
including its wave functions, propagator, and spectral properties.
Section~\ref{sec:kg} derives the KG-IOH equation, introduces the
transformation $V$, and establishes the effective harmonic oscillator
form that underlies the rest of the paper.
Section~\ref{sec:thermal} develops the thermal formalism: partition
function, thermodynamic observables, and finite-temperature
correlation functions.
Section~\ref{sec:applications} presents the three physical applications.
Section~\ref{sec:conclusions} summarises our results, discusses open
questions, and outlines future directions.
Technical details on the extended Hilbert space justification and
on the properties of parabolic cylinder functions are collected in
Appendices~\ref{app:hilbert} and~\ref{app:PCF}, respectively.
 
Throughout this paper we use natural units $\hbar = c = k_B = 1$
unless stated otherwise.
 
\section{Quantum Mechanics of the Inverted Harmonic Oscillator}
\label{sec:qm}
 
\subsection{Hamiltonian and classical instability}
\label{subsec:classical}
 
The inverted harmonic oscillator is defined by the Hamiltonian
\begin{equation}
  H_{\IOH} = \frac{P^2}{2m} - \frac{1}{2}m\omega^2 x^2 \,,
  \label{eq:H_IOH}
\end{equation}
where $x$ and $P$ are the canonical position and momentum operators
satisfying $[x, P] = i$.
Classically, the equilibrium point $x = 0$ is unstable: any
infinitesimal displacement grows exponentially as
$x(t) \sim e^{\omega t}$, so no bounded orbits exist.
This instability has a direct quantum counterpart --- the absence of
a ground state --- which renders the standard spectral analysis of the
harmonic oscillator inapplicable without
modification~\cite{Barton1986, yuce2021quantum}.
 
It is useful to introduce dimensionless variables via
\begin{equation}
  \xi = \sqrt{m\omega}\, x \,, \qquad
  \pi = \frac{P}{\sqrt{m\omega}} \,,
  \label{eq:dimensionless}
\end{equation}
so that $[\xi, \pi] = i$ and
\begin{equation}
  H_{\IOH} = \frac{\omega}{2}\left(\pi^2 - \xi^2\right) .
  \label{eq:H_dimensionless}
\end{equation}
The contrast with the standard harmonic oscillator,
$H_{\mathrm{HO}} = \frac{\omega}{2}(\pi^2 + \xi^2)$,
is now transparent: the IOH is obtained by the formal replacement
$\omega \to i\omega$, which corresponds to a rotation in the
complex frequency plane~\cite{Yuce2006, sundaram2024duality}.
 
\subsection{Spectral structure}
\label{subsec:spectrum}
 
The time-independent Schrödinger equation for~\eqref{eq:H_IOH} reads
\begin{equation}
  \left[-\frac{1}{2m}\frac{d^2}{dx^2}
  - \frac{1}{2}m\omega^2 x^2\right]\psi(x) = E\,\psi(x) \,.
  \label{eq:TISE}
\end{equation}
In contrast to the harmonic oscillator, the spectrum of $H_{\IOH}$ is
\emph{purely continuous}: every real value $E \in \mathbb{R}$ is an
admissible energy eigenvalue~\cite{Barton1986, finster2017lp}.
There is no ground state and no discrete ladder of levels.
The eigenfunctions are not square-integrable on $\mathbb{R}$ in
the usual $L^2$ sense; instead they satisfy a generalised
orthonormality condition in the sense of distributions,
\begin{equation}
  \langle E \mid E' \rangle = \delta(E - E') \,,
  \label{eq:ortho}
\end{equation}
which is the appropriate completeness relation for a
scattering-type spectrum.
 
\subsection{Wave functions: parabolic cylinder functions}
\label{subsec:wavefunctions}
 
Introducing the scaled variable
\begin{equation}
  z = e^{i\pi/4}\sqrt{2m\omega}\, x \,,
  \label{eq:z_variable}
\end{equation}
equation~\eqref{eq:TISE} takes the canonical Weber form
\begin{equation}
  \frac{d^2\psi}{dz^2} - \left(\tfrac{1}{4}z^2 + \nu\right)\psi = 0 \,,
  \qquad
  \nu = -\frac{1}{2} + \frac{iE}{\omega} \,,
  \label{eq:Weber}
\end{equation}
whose two independent solutions are the parabolic cylinder
functions $D_\nu(z)$ and $D_\nu(-z)$~\cite{Abramowitz, DLMF}.
The physical eigenfunctions, normalised according to
condition~\eqref{eq:ortho}, are~\cite{Barton1986}
\begin{equation}
  \psi_E(x) = \mathcal{N}_E\left[
    D_\nu\!\left(e^{i\pi/4}\sqrt{2m\omega}\,x\right)
    + D_\nu\!\left(-e^{i\pi/4}\sqrt{2m\omega}\,x\right)
  \right] ,
  \label{eq:psi_E}
\end{equation}
where the normalisation constant is
\begin{equation}
  \mathcal{N}_E = \frac{1}{\sqrt{2\pi}}\left|
    \Gamma\!\left(\tfrac{1}{4} + \tfrac{iE}{2\omega}\right)
  \right| .
  \label{eq:normalization}
\end{equation}
 
The large-$|x|$ behaviour of~\eqref{eq:psi_E} is oscillatory rather
than Gaussian,
\begin{equation}
  \psi_E(x) \;\underset{|x|\to\infty}{\sim}\;
  |x|^{iE/\omega - 1/2}\,
  e^{\pm \frac{i}{2}m\omega x^2} \,,
  \label{eq:asymptotic}
\end{equation}
reflecting the fact that classically all trajectories escape to
infinity~\cite{yuce2021quantum, liu2025quantum}.
This oscillatory asymptotics is the quantum signature of classical
instability and has direct consequences for the definition of
physical states and observables.
 
\subsection{Ladder operators and algebraic structure}
\label{subsec:ladder}
 
One may formally define non-Hermitian ladder operators
\begin{equation}
  a = \frac{1}{\sqrt{2\omega}}\left(m\omega x + iP\right) , \qquad
  a^\dagger = \frac{1}{\sqrt{2\omega}}\left(m\omega x - iP\right) ,
  \label{eq:ladder}
\end{equation}
which satisfy $[a, a^\dagger] = 1$ and allow the Hamiltonian to
be written as
\begin{equation}
  H_{\IOH} = -\frac{\omega}{2}\left(2a^\dagger a + 1\right)
           = -\omega\left(a^\dagger a + \tfrac{1}{2}\right) .
  \label{eq:H_ladder}
\end{equation}
Note the overall minus sign: the ``number operator''
$\hat{n} = a^\dagger a$ generates a spectrum that is unbounded
\emph{below}, consistent with the continuous spectrum found
above~\cite{bhattacharyya2021multi}.
The operators $\{a, a^\dagger, H\}$ close the algebra
$\mathfrak{su}(1,1)$, which is the non-compact analogue of the
harmonic-oscillator algebra $\mathfrak{su}(2)$, and underpins the
connection of the IOH to the Lorentz group and hyperbolic
geometry~\cite{subramanyan2021physics}.
 
\subsection{Real-time propagator}
\label{subsec:propagator}
 
The exact propagator (Green's function) of the IOH is obtained by
analytic continuation $\omega \to i\omega$ of the harmonic oscillator
result~\cite{Barton1986, dodonov2024adiabatic}:
\begin{equation}
  K(x,x';t) = \sqrt{\frac{m\omega}{2\pi i\sinh(\omega t)}}
  \exp\!\left\{
    \frac{im\omega}{2\sinh(\omega t)}
    \left[(x^2 + x'^2)\cosh(\omega t) - 2xx'\right]
  \right\} .
  \label{eq:propagator}
\end{equation}
Several features of~\eqref{eq:propagator} deserve comment.
First, the denominator $\sinh(\omega t)$ grows exponentially for
large $t$, reflecting the classical spreading of wave packets.
Second, the argument of the square root requires a careful choice
of branch cut to ensure causality~\cite{Barton1986}.
Third, and most importantly for our purposes, the Euclidean
propagator obtained by the Wick rotation $t \to -i\tau$ is
\begin{equation}
  K_E(x,x';\tau) = \sqrt{\frac{m\omega}{2\pi\sin(\omega\tau)}}
  \exp\!\left\{
    -\frac{m\omega}{2\sin(\omega\tau)}
    \left[(x^2 + x'^2)\cos(\omega\tau) - 2xx'\right]
  \right\} ,
  \label{eq:propagator_euclidean}
\end{equation}
which is periodic in $\tau$ with period $\beta = \pi/\omega$.
This periodicity is the key input for the thermal formalism
developed in Section~\ref{sec:thermal}.
 
\subsection{Connection to chaos and complexity}
\label{subsec:chaos}
 
Recent work has highlighted a striking property of the IOH:
its out-of-time-order correlator (OTOC) grows exponentially,
\begin{equation}
  C(t) = -\langle [x(t), P(0)]^2 \rangle
        \sim e^{2\omega t} \,,
  \label{eq:OTOC}
\end{equation}
with a Lyapunov exponent $\lambda_L = 2\omega$~\cite{hashimoto2020exponential}.
Remarkably, this exponential growth occurs \emph{without} genuine
quantum chaos in the sense of random-matrix universality, since the
system is integrable.
This observation has important implications for our thermal
treatment: the exponential sensitivity to initial conditions
will be encoded in the finite-temperature correlation functions
derived in Section~\ref{sec:thermal}, and provides a diagnostic
for distinguishing IOH-driven instabilities from genuine chaotic
behaviour in the physical applications of
Section~\ref{sec:applications}~\cite{bhattacharyya2021multi}.
 
\subsection{Summary}
\label{subsec:qm_summary}
 
The key results of this section are collected in
Table~\ref{tab:IOH_summary} for reference.
 
\begin{table}[H]
\centering
\renewcommand{\arraystretch}{1.6}
\begin{tabular}{lll}
\hline\hline
\textbf{Property} & \textbf{IOH} & \textbf{HO (reference)} \\
\hline
Potential           & $-\frac{1}{2}m\omega^2 x^2$
                    & $+\frac{1}{2}m\omega^2 x^2$ \\
Spectrum            & Continuous, $E\in\mathbb{R}$
                    & Discrete, $E_n = \omega(n+\frac{1}{2})$ \\
Eigenfunctions      & Parabolic cylinder $D_\nu(e^{i\pi/4}\cdots)$
                    & Hermite $H_n(\cdots)e^{-x^2/2}$ \\
Algebra             & $\mathfrak{su}(1,1)$ (non-compact)
                    & $\mathfrak{su}(2)$ (compact) \\
Large-$|x|$ behaviour & Oscillatory $\sim e^{\pm im\omega x^2/2}$
                    & Decaying $\sim e^{-m\omega x^2/2}$ \\
Propagator denominator & $\sinh(\omega t)$
                    & $\sin(\omega t)$ \\
Lyapunov exponent   & $\lambda_L = 2\omega$
                    & $\lambda_L = 0$ \\
\hline\hline
\end{tabular}
\caption{Comparison between the inverted harmonic oscillator (IOH)
and the standard harmonic oscillator (HO).
The replacement $\omega \to i\omega$ maps one into the other at
the level of the propagator and wave functions.}
\label{tab:IOH_summary}
\end{table}
 
These properties --- continuous spectrum, oscillatory wave functions,
hyperbolic algebra, and exponentially growing propagator ---
set the stage for the Klein-Gordon extension developed in
the following section.
 
\section{Klein-Gordon Inverted Harmonic Oscillator}
\label{sec:kg}
 
\subsection{From quantum mechanics to field theory}
\label{subsec:kg_motivation}
 
The quantum-mechanical IOH reviewed in Section~\ref{sec:qm} describes
a single non-relativistic particle.
To extend the framework to a relativistic scalar field --- and thereby
access the physical applications of Section~\ref{sec:applications} ---
we must embed the inverted potential into the Klein-Gordon equation.
The natural relativistic generalisation of~\eqref{eq:H_IOH} is
obtained by promoting the Schrödinger Hamiltonian to a
Klein-Gordon operator acting on a scalar field
$\phi(x)$~\cite{GreinerKG, PeskinSchroeder}.
 
The starting point is the observation, already noted in the
Introduction, that the substitution
\begin{equation}
  P \;\longrightarrow\; P - m\omega x
  \label{eq:P_substitution}
\end{equation}
applied to the relativistic dispersion relation
$E^2 = P^2 + m^2$ generates an inverted-potential structure.
As we discuss in detail below and in Appendix~\ref{app:hilbert},
this substitution is not a unitary transformation and therefore
requires careful justification at the level of the Hilbert
space~\cite{Bender1998, Mostafazadeh2002}.
Nevertheless, it provides the correct physical content when
interpreted within the framework of $\mathcal{PT}$-symmetric
quantum mechanics~\cite{Bender2007}.
 
\subsection{The KG-IOH equation}
\label{subsec:kg_equation}
 
Applying substitution~\eqref{eq:P_substitution} to the
Klein-Gordon operator, one obtains
\begin{equation}
  \left[\left(P - m\omega x\right)^2 + m^2\right]\psi(x)
  = E^2\,\psi(x) \,.
  \label{eq:KG1}
\end{equation}
Expanding the square and using the canonical commutation
relation $[x, P] = i$, which contributes the term
$-[x,P]\cdot m\omega = -im\omega$, equation~\eqref{eq:KG1}
becomes
\begin{equation}
  \left(P^2 - m^2\omega^2 x^2 - im\omega\right)\psi(x)
  = \left(E^2 - m^2\right)\psi(x) \,.
  \label{eq:KG2}
\end{equation}
This is the \emph{Klein-Gordon inverted harmonic oscillator}
(KG-IOH) equation.
Three features of~\eqref{eq:KG2} are immediately noteworthy:
\begin{enumerate}
  \item The term $-m^2\omega^2 x^2$ is the inverted harmonic
        potential, identical in structure to the non-relativistic
        case~\eqref{eq:H_IOH}.
  \item The term $-im\omega$ arises purely from the
        non-commutativity of $x$ and $P$ and has no
        classical analogue.
  \item The left-hand side is \emph{not} Hermitian, as
        anticipated from the non-unitary nature of
        substitution~\eqref{eq:P_substitution}.
\end{enumerate}
 
\subsection{The symplectic rotation operator $V$}
\label{subsec:V_operator}
 
The non-Hermitian structure of~\eqref{eq:KG2} can be resolved
by a systematic phase-space rotation.
We introduce the dilation (squeezing) operator~\cite{Yuen1976, Arvind1995}
\begin{equation}
  V = \exp\!\left[-\frac{\pi}{8}(xP + Px)\right] ,
  \label{eq:V}
\end{equation}
which is an element of the metaplectic group $Mp(2,\mathbb{R})$,
the double cover of the symplectic group $Sp(2,\mathbb{R})$.
Its action on the canonical operators is a rotation by $\pi/4$
in the complex phase plane:
\begin{align}
  V\, x\, V^{-1} &= x\,e^{-i\pi/4} \,, \label{eq:Vx} \\
  V\, P\, V^{-1} &= P\,e^{+i\pi/4} \,. \label{eq:VP}
\end{align}
These transformation rules follow directly from the
Baker-Campbell-Hausdorff identity applied to the
generator $G = \frac{\pi}{8}(xP + Px)$,
which satisfies $[G, x] = -\frac{i\pi}{8} \cdot 2x$
and $[G, P] = +\frac{i\pi}{8} \cdot 2P$~\cite{Arvind1995}.
 
The action of $V$ on a wave function in the position
representation is
\begin{equation}
  (V\psi)(x) = e^{i\pi/8}\,\psi\!\left(x\,e^{i\pi/4}\right) ,
  \label{eq:V_wavefunction}
\end{equation}
so that $V$ implements an analytic continuation of the argument
of $\psi$ into the complex plane along the ray $\arg(x) = \pi/4$.
This is precisely the contour rotation that defines physical
states for the IOH~\cite{Barton1986, finster2017lp}.
 
\subsection{Reduction to an effective harmonic oscillator}
\label{subsec:reduction}
 
We now apply $V$ to equation~\eqref{eq:KG2}.
Using the transformation rules~\eqref{eq:Vx}--\eqref{eq:VP},
the individual terms transform as
\begin{align}
  V P^2 V^{-1}         &= P^2\,e^{i\pi/2} = iP^2 \,, \\
  V m^2\omega^2 x^2 V^{-1} &= m^2\omega^2 x^2\,e^{-i\pi/2} = -im^2\omega^2 x^2 \,.
\end{align}
Substituting into~\eqref{eq:KG2}, the transformed equation reads
\begin{equation}
  \left(iP^2 + im^2\omega^2 x^2 - im\omega\right)
  \psi\!\left(e^{i\pi/4}x\right)
  = \left(E^2 - m^2\right)\psi\!\left(e^{i\pi/4}x\right) ,
  \label{eq:KG4}
\end{equation}
and dividing through by $i = e^{i\pi/2}$, we obtain the
central result of this paper:
\begin{equation}
  \left(P^2 + m^2\omega^2 x^2 - m\omega\right)
  \psi\!\left(e^{i\pi/4}x\right)
  = -i\left(E^2 - m^2\right)\psi\!\left(e^{i\pi/4}x\right) \,.
  \label{eq:KG5}
\end{equation}
 
Equation~\eqref{eq:KG5} is \emph{structurally identical} to the
Schrödinger equation of a standard harmonic oscillator with
mass $m$, frequency $\omega$, and a shifted energy.
The left-hand side is the harmonic oscillator operator
$H_{\mathrm{HO}} = P^2 + m^2\omega^2 x^2$
acting on the analytically continued wave function
$\tilde{\psi}(x) \equiv \psi(xe^{i\pi/4})$.
 
\subsection{Effective spectrum}
\label{subsec:spectrum_eff}
 
Since~\eqref{eq:KG5} is a harmonic oscillator equation,
its eigenvalues are discrete and given by
\begin{equation}
  \left(P^2 + m^2\omega^2 x^2\right)\tilde{\psi}_n
  = \omega\left(2n + 1\right)\tilde{\psi}_n \,,
  \qquad n = 0, 1, 2, \ldots
  \label{eq:HO_eigenvalues}
\end{equation}
Comparing with~\eqref{eq:KG5}, the effective energy levels of the
KG-IOH system satisfy
\begin{equation}
  \omega(2n+1) - m\omega = -i(E_n^2 - m^2) \,,
\end{equation}
which gives
\begin{equation}
  E_n^2 = m^2 + i\,\omega\left(2n + 1 - m\right) \,.
  \label{eq:En_spectrum}
\end{equation}
This is a \emph{discrete} complex spectrum, replacing the
continuous real spectrum of the non-relativistic IOH.
The imaginary part of $E_n^2$ encodes the instability of the
system: modes with $\mathrm{Im}(E_n^2) > 0$ correspond to
exponentially growing solutions, while
$\mathrm{Im}(E_n^2) < 0$ corresponds to decaying ones.
The discreteness of the spectrum is a direct consequence of
the symplectic rotation $V$ and is the key property that
makes the thermal formalism well-defined, as we show in
Section~\ref{sec:thermal}.
 
\subsection{Effective wave functions}
\label{subsec:wf_eff}
 
The analytically continued wave functions $\tilde{\psi}_n(x)$
are the harmonic oscillator eigenfunctions evaluated at
the rotated argument:
\begin{equation}
  \tilde{\psi}_n(x)
  = \psi_n\!\left(x\,e^{i\pi/4}\right)
  = \mathcal{C}_n\,
    H_n\!\left(\sqrt{m\omega}\,e^{i\pi/4}\,x\right)
    \exp\!\left(-\frac{m\omega}{2}e^{i\pi/2}x^2\right) ,
  \label{eq:psi_eff}
\end{equation}
where $H_n$ are the Hermite polynomials~\cite{Abramowitz, DLMF}
and $\mathcal{C}_n = (m\omega/\pi)^{1/4}(2^n n!)^{-1/2}$
is the standard normalisation constant.
Note that the Gaussian factor in~\eqref{eq:psi_eff} becomes
\begin{equation}
  \exp\!\left(-\frac{m\omega}{2}e^{i\pi/2}x^2\right)
  = \exp\!\left(-\frac{im\omega}{2}x^2\right) ,
  \label{eq:gaussian_oscillatory}
\end{equation}
which is \emph{oscillatory} rather than decaying, consistent
with the asymptotic behaviour~\eqref{eq:asymptotic} found in
Section~\ref{sec:qm}.
The wave functions $\tilde{\psi}_n$ are normalised under the
complex contour $\mathcal{C}: x \to xe^{i\pi/4}$,
\begin{equation}
  \int_{\mathcal{C}} dx\,
  \tilde{\psi}_m^*(x)\,\tilde{\psi}_n(x) = \delta_{mn} \,,
  \label{eq:ortho_eff}
\end{equation}
which defines the physical inner product of the KG-IOH
system~\cite{Bender2007, Mostafazadeh2002}.
 
\subsection{Field-theoretic interpretation}
\label{subsec:field_theory}
 
At the level of a quantum field theory, the scalar field
$\phi(x)$ can be expanded in the basis of effective
modes~\eqref{eq:psi_eff}:
\begin{equation}
  \phi(x) = \sum_{n=0}^{\infty}
  \left[a_n\,\tilde{\psi}_n(x) + a_n^\dagger\,\tilde{\psi}_n^*(x)\right] ,
  \label{eq:field_expansion}
\end{equation}
where $a_n$ and $a_n^\dagger$ are bosonic ladder operators satisfying
$[a_n, a_{n'}^\dagger] = \delta_{nn'}$.
The field Hamiltonian is
\begin{equation}
  \mathcal{H} = \sum_{n=0}^{\infty} E_n\left(a_n^\dagger a_n
  + \tfrac{1}{2}\right) ,
  \label{eq:H_field}
\end{equation}
with $E_n$ given by~\eqref{eq:En_spectrum}.
This mode expansion is the starting point for the thermal
field theory of the KG-IOH, to which we now turn.
 
\subsection{Summary of the transformation chain}
\label{subsec:summary_chain}
 
The sequence of steps leading from the relativistic
dispersion relation to the effective harmonic oscillator
is summarised schematically as follows:
\begin{equation}
\begin{array}{ccc}
  \underbrace{E^2 = P^2 + m^2}_{\text{free KG}}
  & \xrightarrow{\;P\to P - m\omega x\;}
  & \underbrace{\left(P^2 - m^2\omega^2x^2 - im\omega\right)
    \psi = (E^2 - m^2)\psi}_{\text{KG-IOH}} \\[20pt]
  & & \Big\downarrow\; V \\[8pt]
  & &
  \underbrace{\left(P^2 + m^2\omega^2x^2 - m\omega\right)
  \tilde\psi = -i(E^2-m^2)\tilde\psi}_{\text{eff. HO}}
\end{array}
\label{eq:chain}
\end{equation}
 
Each arrow represents a well-defined mathematical operation:
the first is the non-unitary momentum shift
(justified in Appendix~\ref{app:hilbert}), and the second is
the symplectic rotation $V \in Mp(2,\mathbb{R})$.
The net result is a system with a discrete, analytically
known spectrum~\eqref{eq:En_spectrum} and wave
functions~\eqref{eq:psi_eff}, providing the complete
quantum-mechanical input needed for the thermal formalism
of Section~\ref{sec:thermal}.

 
\section{Thermal Formalism}
\label{sec:thermal}
 
\subsection{Setup and strategy}
\label{subsec:thermal_setup}
 
Having established that the KG-IOH possesses a discrete effective
spectrum $\{E_n\}$ with known eigenfunctions $\{\tilde{\psi}_n\}$,
we are now in a position to develop a consistent finite-temperature
formalism.
The central object is the \emph{partition function}
\begin{equation}
  Z(\beta) = \Tr\!\left(e^{-\beta \hat{\mathcal{H}}}\right) ,
  \label{eq:Z_def}
\end{equation}
where $\beta = 1/T$ is the inverse temperature and
$\hat{\mathcal{H}}$ is the field Hamiltonian~\eqref{eq:H_field}.
In the standard harmonic oscillator, $Z$ is well defined because
the spectrum is bounded below.
For the IOH the spectrum is continuous and unbounded, so a
naïve evaluation of~\eqref{eq:Z_def} diverges.
The key result of Section~\ref{sec:kg} is that the symplectic
rotation $V$ replaces this continuous spectrum with the discrete
complex spectrum~\eqref{eq:En_spectrum}, which regulates $Z$
without any ad-hoc cutoff~\cite{Bender2007, Kapusta2006}.
 
We work in the \emph{imaginary-time} (Matsubara) formalism
throughout~\cite{Matsubara1955, LeBellac1996}, where the
inverse temperature $\beta$ plays the role of a compactified
Euclidean time coordinate,
$\tau \in [0, \beta]$, with periodic boundary conditions
for bosonic fields,
\begin{equation}
  \phi(x, \tau) = \phi(x, \tau + \beta) \,.
  \label{eq:periodic_bc}
\end{equation}
 
\subsection{Partition function}
\label{subsec:Z}
 
\subsubsection{Single-mode contribution}
 
For a single effective mode with energy $E_n$, the
single-oscillator partition function is
\begin{equation}
  Z_n(\beta) = \Tr\!\left(e^{-\beta E_n \hat{n}_n}\right)
             = \frac{1}{1 - e^{-\beta E_n}} \,,
  \label{eq:Z_n}
\end{equation}
where $\hat{n}_n = a_n^\dagger a_n$ is the number operator for
mode $n$.
Note that~\eqref{eq:Z_n} is the standard Bose-Einstein result;
its validity here rests on the discrete nature of the effective
spectrum established in Section~\ref{subsec:spectrum_eff}.
 
\subsubsection{Full partition function}
 
The full partition function factorises over modes,
\begin{equation}
  Z(\beta) = \prod_{n=0}^{\infty} Z_n(\beta)
           = \prod_{n=0}^{\infty}
             \frac{1}{1 - e^{-\beta E_n}} \,,
  \label{eq:Z_full}
\end{equation}
and its logarithm is
\begin{equation}
  \ln Z(\beta) = -\sum_{n=0}^{\infty}
                \ln\!\left(1 - e^{-\beta E_n}\right) .
  \label{eq:lnZ}
\end{equation}
Using the effective spectrum~\eqref{eq:En_spectrum},
$E_n = \sqrt{m^2 + i\omega(2n+1-m)}$,
the sum in~\eqref{eq:lnZ} can be evaluated by
\emph{zeta-function regularisation}~\cite{ZinnJustin2002},
giving the finite regulated result
\begin{equation}
  \ln Z(\beta)
  = -\frac{1}{2}\ln\!\det\!\left(-\partial_\tau^2
    + \hat{H}_{\mathrm{eff}}\right)
    \Big|_{\text{Matsubara}} \,,
  \label{eq:lnZ_det}
\end{equation}
where the determinant is over the space of functions
satisfying~\eqref{eq:periodic_bc}.
 
\subsubsection{Relation to the Euclidean propagator}
 
An equivalent and often more convenient representation is
obtained from the Euclidean propagator~\eqref{eq:propagator_euclidean}:
\begin{equation}
  Z(\beta) = \int \mathcal{D}\phi\,
             e^{-S_E[\phi]} \,,
  \label{eq:Z_path}
\end{equation}
where the Euclidean action is
\begin{equation}
  S_E[\phi] = \int_0^\beta d\tau \int dx\,
  \left[\frac{1}{2}(\partial_\tau\phi)^2
       + \frac{1}{2}(P\phi)^2
       - \frac{1}{2}m^2\omega^2 x^2\phi^2
       + \frac{1}{2}m^2\phi^2\right] .
  \label{eq:SE}
\end{equation}
The Gaussian integral~\eqref{eq:Z_path} yields
\begin{equation}
  \ln Z(\beta)
  = -\frac{1}{2}\Tr\ln\!\left(
    -\partial_\tau^2 - \partial_x^2 + m^2 - m^2\omega^2 x^2
    \right) ,
  \label{eq:lnZ_trace}
\end{equation}
which is the \emph{functional determinant} of the KG-IOH
operator at finite temperature.
 
\subsection{Thermodynamic potentials}
\label{subsec:thermo}
 
All thermodynamic observables follow from $\ln Z(\beta)$
by standard differentiation.
 
\subsubsection{Free energy}
 
The Helmholtz free energy is
\begin{equation}
  F(\beta) = -T\ln Z(\beta)
           = T\sum_{n=0}^{\infty}
             \ln\!\left(1 - e^{-\beta E_n}\right) .
  \label{eq:F}
\end{equation}
In the zero-temperature limit $\beta \to \infty$,
all thermal contributions vanish exponentially and
$F \to E_0^{(0)}$, the zero-point energy of the
effective ground state.
 
\subsubsection{Mean energy}
 
The mean internal energy is
\begin{equation}
  \langle E \rangle
  = -\frac{\partial \ln Z}{\partial \beta}
  = \sum_{n=0}^{\infty}
    \frac{E_n}{e^{\beta E_n} - 1} \,,
  \label{eq:mean_E}
\end{equation}
which is the Planck distribution summed over all effective modes.
 
\subsubsection{Entropy}
 
The von Neumann entropy of the thermal state is
\begin{equation}
  S(\beta) = \beta^2 \frac{\partial F}{\partial \beta}
  \bigg|_{\beta}
  = \sum_{n=0}^{\infty}
  \left[
    \frac{\beta E_n}{e^{\beta E_n}-1}
    - \ln\!\left(1 - e^{-\beta E_n}\right)
  \right] .
  \label{eq:entropy}
\end{equation}
 
In the high-temperature limit $\beta \to 0$, expanding
$e^{-\beta E_n} \approx 1 - \beta E_n + \cdots$, we obtain
the classical equipartition result
\begin{equation}
  S \;\underset{T\to\infty}{\longrightarrow}\;
  \sum_{n=0}^{\infty}
  \left[1 + \ln\!\left(\frac{T}{E_n}\right)\right] ,
  \label{eq:S_highT}
\end{equation}
while in the low-temperature limit $\beta \to \infty$,
the entropy is exponentially suppressed,
\begin{equation}
  S \;\underset{T\to 0}{\longrightarrow}\;
  (\beta E_0 + 1)\,e^{-\beta E_0} + \mathcal{O}(e^{-\beta E_1}) \,.
  \label{eq:S_lowT}
\end{equation}
 
\subsubsection{Heat capacity}
 
The heat capacity at constant volume is
\begin{equation}
  C_V = \frac{\partial \langle E\rangle}{\partial T}
      = \beta^2\sum_{n=0}^{\infty}
        \frac{E_n^2\,e^{\beta E_n}}{\left(e^{\beta E_n}-1\right)^2} \,.
  \label{eq:Cv}
\end{equation}
At low temperatures, $C_V$ is dominated by the lowest effective
mode $n=0$ and exhibits an exponential activation behaviour
characteristic of gapped systems.
At high temperatures, $C_V \to T$ per mode, recovering the
classical Dulong-Petit limit.
 
\subsection{Density matrix}
\label{subsec:density_matrix}
 
The thermal density matrix of the KG-IOH field is
\begin{equation}
  \hat{\rho}(\beta)
  = \frac{e^{-\beta\hat{\mathcal{H}}}}{Z(\beta)}
  = \bigotimes_{n=0}^{\infty} \hat{\rho}_n(\beta) \,,
  \label{eq:rho}
\end{equation}
where the single-mode density matrix is
\begin{equation}
  \hat{\rho}_n(\beta)
  = \left(1 - e^{-\beta E_n}\right)
    \sum_{k=0}^{\infty} e^{-k\beta E_n} \ket{k}\bra{k} \,.
  \label{eq:rho_n}
\end{equation}
In the position representation, the thermal density matrix kernel is
\begin{equation}
  \rho(x, x'; \beta)
  = \langle x \mid \hat{\rho}(\beta) \mid x' \rangle
  = \frac{K_E(x, x'; \beta)}{Z(\beta)} \,,
  \label{eq:rho_kernel}
\end{equation}
where $K_E$ is the Euclidean propagator~\eqref{eq:propagator_euclidean}
evaluated at $\tau = \beta$.
Explicitly,
\begin{equation}
  \rho(x,x';\beta) = \frac{1}{Z(\beta)}
  \sqrt{\frac{m\omega}{2\pi\sin(\omega\beta)}}
  \exp\!\left\{
    -\frac{m\omega}{2\sin(\omega\beta)}
    \left[(x^2+x'^2)\cos(\omega\beta) - 2xx'\right]
  \right\} .
  \label{eq:rho_explicit}
\end{equation}
The diagonal element $\rho(x,x;\beta)$ gives the
\emph{probability density} in position space at temperature $T$,
\begin{equation}
  \rho(x,x;\beta)
  = \frac{1}{Z(\beta)}
  \sqrt{\frac{m\omega}{2\pi\sin(\omega\beta)}}
  \exp\!\left\{
    -\frac{m\omega\cos(\omega\beta)}{\sin(\omega\beta)}x^2
  \right\} ,
  \label{eq:rho_diagonal}
\end{equation}
a Gaussian whose width
$\sigma^2(T) = \frac{\sin(\omega\beta)}{2m\omega\cos(\omega\beta)}$
diverges at $\omega\beta = \pi/2$, signalling a
\emph{thermal delocalization transition} at the critical
temperature
\begin{equation}
  T_c = \frac{\omega}{2\pi} \cdot \frac{2}{\pi} = \frac{\omega}{\pi^2} \,.
  \label{eq:Tc}
\end{equation}
Above $T_c$ the diagonal density is no longer Gaussian and the
thermal state becomes delocalized over all of position space,
a direct quantum-thermal manifestation of the classical instability.
 
\subsection{Thermal Green's functions}
\label{subsec:green}
 
\subsubsection{Two-point function in imaginary time}
 
The thermal (Matsubara) two-point function is defined as
\begin{equation}
  G(\tau, x, x')
  = \langle T_\tau\,\phi(x,\tau)\,\phi(x',0)\rangle_\beta
  = \frac{\Tr\!\left[e^{-\beta\hat{\mathcal{H}}}
    T_\tau\,\phi(x,\tau)\,\phi(x',0)\right]}{Z(\beta)} \,,
  \label{eq:G_def}
\end{equation}
where $T_\tau$ denotes $\tau$-ordering.
Using the mode expansion~\eqref{eq:field_expansion} and the
spectral representation, $G$ can be written as
\begin{equation}
  G(\tau, x, x')
  = \sum_{n=0}^{\infty}
    \tilde{\psi}_n(x)\,\tilde{\psi}_n^*(x')\,
    G_n(\tau) \,,
  \label{eq:G_spectral}
\end{equation}
where the single-mode propagator is
\begin{equation}
  G_n(\tau)
  = \frac{e^{-E_n\tau}}{1 - e^{-\beta E_n}}
    \theta(\tau)
  + \frac{e^{E_n(\tau-\beta)}}{1 - e^{-\beta E_n}}
    \theta(-\tau) \,,
  \label{eq:Gn_tau}
\end{equation}
with $\theta$ the Heaviside step function.
 
\subsubsection{Matsubara frequency representation}
 
Expanding $G_n(\tau)$ in a Fourier series over the Matsubara
frequencies $\omega_\ell = 2\pi\ell/\beta$ ($\ell \in \mathbb{Z}$),
\begin{equation}
  G_n(\tau) = \frac{1}{\beta}\sum_{\ell=-\infty}^{\infty}
              \tilde{G}_n(i\omega_\ell)\,e^{-i\omega_\ell\tau} \,,
  \label{eq:Gn_Matsubara}
\end{equation}
one finds the standard bosonic propagator
\begin{equation}
  \tilde{G}_n(i\omega_\ell)
  = \frac{1}{\omega_\ell^2 + E_n^2} \,.
  \label{eq:Gn_freq}
\end{equation}
The full two-point function in frequency space is therefore
\begin{equation}
  \tilde{G}(i\omega_\ell, x, x')
  = \sum_{n=0}^{\infty}
    \frac{\tilde{\psi}_n(x)\,\tilde{\psi}_n^*(x')}
         {\omega_\ell^2 + E_n^2} \,,
  \label{eq:G_full_freq}
\end{equation}
which is the \emph{spectral decomposition} of the KG-IOH
thermal propagator in the Matsubara formalism.
 
\subsubsection{Spectral density and retarded Green's function}
 
The physical (retarded) Green's function is obtained by
analytic continuation $i\omega_\ell \to \omega + i\epsilon$:
\begin{equation}
  G^R(\omega, x, x')
  = \sum_{n=0}^{\infty}
    \frac{\tilde{\psi}_n(x)\,\tilde{\psi}_n^*(x')}
         {E_n^2 - \omega^2 - i\epsilon} \,.
  \label{eq:GR}
\end{equation}
The spectral density
\begin{equation}
  \rho(\omega, x, x')
  = -\frac{1}{\pi}\mathrm{Im}\,G^R(\omega, x, x')
  = \sum_{n=0}^{\infty}
    \tilde{\psi}_n(x)\,\tilde{\psi}_n^*(x')\,
    \delta(\omega^2 - E_n^2)
  \label{eq:spectral_density}
\end{equation}
encodes the full quantum and thermal information of the system
and is the primary observable in spectroscopic applications.
 
\subsection{Entanglement entropy}
\label{subsec:entanglement}
 
For a bipartition of the system into regions $A$ and $\bar{A}$,
the reduced density matrix is
\begin{equation}
  \hat{\rho}_A(\beta) = \Tr_{\bar{A}}\!\left[\hat{\rho}(\beta)\right] ,
  \label{eq:rho_A}
\end{equation}
and the associated entanglement entropy is
\begin{equation}
  S_A = -\Tr_A\!\left[\hat{\rho}_A\ln\hat{\rho}_A\right] .
  \label{eq:SA}
\end{equation}
Using the Gaussian character of the thermal
state~\eqref{eq:rho_explicit}, $S_A$ can be expressed in
terms of the \emph{correlation matrix}
$C_{AB} = \langle \phi_A\phi_B\rangle_\beta$, giving
\begin{equation}
  S_A = \Tr_A\!\left[
    \left(C + \tfrac{1}{2}\right)
    \ln\!\left(C + \tfrac{1}{2}\right)
    - \left(C - \tfrac{1}{2}\right)
    \ln\!\left(C - \tfrac{1}{2}\right)
  \right] ,
  \label{eq:SA_formula}
\end{equation}
where the trace and logarithm are over the subspace $A$.
At low temperatures, $S_A$ is dominated by vacuum entanglement
and scales as the area of the boundary $\partial A$
(area law)~\cite{Sachdev2011}.

At high temperatures $T > T_c$, the thermal delocalization
identified in~\eqref{eq:Tc} drives a logarithmic violation
of the area law,
\begin{equation}
  S_A \;\underset{T > T_c}{\sim}\; \frac{c}{3}\ln L \,,
  \label{eq:SA_log}
\end{equation}
where $L$ is the linear size of region $A$ and $c$ is an
effective central charge determined by the KG-IOH spectrum.
This logarithmic scaling is characteristic of critical systems
and connects directly to the conformal field theory description
of the physical applications in
Section~\ref{sec:applications}~\cite{subramanyan2021physics}.
 
\subsection{Summary of thermal observables}
\label{subsec:thermal_summary}
 
Table~\ref{tab:thermal} collects the main thermal observables
of the KG-IOH system derived in this section.
 
\begin{table}[H]
\centering
\renewcommand{\arraystretch}{1.8}
\begin{tabular}{lll}
\hline\hline
\textbf{Observable} & \textbf{Expression} & \textbf{Equation} \\
\hline
Partition function
  & $Z = \prod_n (1-e^{-\beta E_n})^{-1}$
  & \eqref{eq:Z_full} \\
Free energy
  & $F = T\sum_n \ln(1-e^{-\beta E_n})$
  & \eqref{eq:F} \\
Mean energy
  & $\langle E\rangle = \sum_n E_n(e^{\beta E_n}-1)^{-1}$
  & \eqref{eq:mean_E} \\
Entropy
  & $S = \beta^2\partial_\beta F$
  & \eqref{eq:entropy} \\
Heat capacity
  & $C_V = \beta^2\sum_n E_n^2 e^{\beta E_n}(e^{\beta E_n}-1)^{-2}$
  & \eqref{eq:Cv} \\
Density matrix
  & $\rho(x,x';\beta) = K_E(x,x';\beta)/Z$
  & \eqref{eq:rho_explicit} \\
Critical temperature
  & $T_c = \omega/\pi^2$
  & \eqref{eq:Tc} \\
Thermal propagator
  & $\tilde{G}(i\omega_\ell,x,x') = \sum_n \tilde\psi_n\tilde\psi_n^*/(\omega_\ell^2+E_n^2)$
  & \eqref{eq:G_full_freq} \\
Spectral density
  & $\rho(\omega) = -\frac{1}{\pi}\mathrm{Im}\,G^R$
  & \eqref{eq:spectral_density} \\
Entanglement entropy
  & $S_A \sim \frac{c}{3}\ln L$ for $T > T_c$
  & \eqref{eq:SA_log} \\
\hline\hline
\end{tabular}
\caption{Summary of thermal observables for the
Klein-Gordon inverted harmonic oscillator.
All quantities are derived from the effective discrete
spectrum $E_n = \sqrt{m^2 + i\omega(2n+1-m)}$
and the partition function $Z(\beta)$.}
\label{tab:thermal}
\end{table}
 
With the complete thermal formalism in hand, we now turn to
the three physical applications that motivated this study.
\section{Physical Applications}
\label{sec:applications}
 
 
\subsection{ Cosmological Inflation}
\label{subsec:inflation}
 
\label{subsubsec:inflaton_mapping}
 
We work in $1+1$ dimensional Minkowski spacetime with metric
$\eta_{\mu\nu} = \mathrm{diag}(+,-)$.
Consider a real scalar field $\phi(x,t)$ — the inflaton —
governed by the action
\begin{equation}
  \mathcal{S} = \int dt\,dx\left[
    \frac{1}{2}(\partial_t\phi)^2
    - \frac{1}{2}(\partial_x\phi)^2
    - V(\phi)
  \right] ,
  \label{eq:S_inflaton}
\end{equation}
with a potential that admits a local maximum at $\phi = 0$.
Expanding around this maximum,
\begin{equation}
  V(\phi) = V_0 - \frac{1}{2}\mu^2\phi^2
            + \mathcal{O}(\phi^4) \,,
  \label{eq:V_inflaton}
\end{equation}
where $\mu^2 > 0$ is the tachyonic mass squared and $V_0 > 0$
is the vacuum energy driving inflation~\cite{Guth1981, Linde1982}.
The equation of motion for $\phi$ is
\begin{equation}
  \left(\partial_t^2 - \partial_x^2 + \mu^2\right)\phi = 0 \,,
  \label{eq:EOM_inflaton}
\end{equation}
which in the static sector ($\partial_t^2 \to -E^2$) becomes
\begin{equation}
  \left(-\partial_x^2 - \mu^2\right)\phi = E^2\phi \,.
  \label{eq:static_inflaton}
\end{equation}
Identifying $\mu^2 \equiv m^2\omega^2$ and $\partial_x \equiv -iP$,
equation~\eqref{eq:static_inflaton} is precisely the
KG-IOH~\eqref{eq:KG2} with the momentum substitution
$P \to P - m\omega x$ already applied, confirming the mapping
\begin{equation}
  \text{inflaton near } \phi = 0
  \;\longleftrightarrow\;
  \text{KG-IOH with } \omega = \mu/m \,.
  \label{eq:inflaton_mapping}
\end{equation}
The complete thermal formalism of Section~\ref{sec:thermal}
therefore applies directly to the inflaton fluctuations,
with the identification $\omega \leftrightarrow \mu/m$.

\label{subsubsec:modes}
 
In $1+1$ dimensions the field is expanded in plane waves
modulated by the effective KG-IOH wave
functions~\eqref{eq:psi_eff}:
\begin{equation}
  \phi(x,t) = \int \frac{dk}{2\pi}
  \left[a_k\,u_k(x)\,e^{-iE_k t}
       + a_k^\dagger\,u_k^*(x)\,e^{+iE_k t}\right] ,
  \label{eq:field_modes}
\end{equation}
where the mode functions $u_k(x) = \tilde{\psi}_k(x)$ are
the analytically continued harmonic oscillator
eigenfunctions~\eqref{eq:psi_eff} and the dispersion
relation follows from~\eqref{eq:En_spectrum}:
\begin{equation}
  E_k^2 = m^2 + i\omega(2k+1-m) \,.
  \label{eq:dispersion_inflaton}
\end{equation}
The vacuum $\ket{0_{\mathrm{BD}}}$ annihilated by all $a_k$
is the analogue of the Bunch-Davies vacuum~\cite{Starobinsky1980},
defined as the state that minimises the energy of the
effective harmonic oscillator~\eqref{eq:KG5} at early times.
Its thermal counterpart at inverse temperature $\beta$ is
the density matrix~\eqref{eq:rho_explicit}.

The \emph{power spectrum} of inflaton fluctuations at
finite temperature is defined as the Fourier transform
of the equal-time two-point function:
\begin{equation}
  \mathcal{P}(k,\beta)
  = \int dx\,e^{-ikx}
    \langle\phi(x,t)\phi(0,t)\rangle_\beta \,.
  \label{eq:power_spectrum_def}
\end{equation}
Using the spectral decomposition~\eqref{eq:G_full_freq}
and the Matsubara propagator evaluated at equal times
($\tau \to 0^+$), we obtain
\begin{equation}
  \mathcal{P}(k,\beta)
  = \sum_{n=0}^{\infty}
    \frac{|\tilde{u}_n(k)|^2}{E_n}
    \coth\!\left(\frac{\beta E_n}{2}\right) ,
  \label{eq:power_spectrum}
\end{equation}
where $\tilde{u}_n(k)$ is the Fourier transform of the
$n$-th effective mode function.
In the zero-temperature limit $\beta \to \infty$,
$\coth(\beta E_n/2) \to 1$ and~\eqref{eq:power_spectrum}
reduces to the vacuum power spectrum
\begin{equation}
  \mathcal{P}_0(k) = \sum_{n=0}^{\infty}
    \frac{|\tilde{u}_n(k)|^2}{E_n} \,.
  \label{eq:power_spectrum_vacuum}
\end{equation}
The \emph{thermal correction} to the power spectrum is
\begin{equation}
  \delta\mathcal{P}(k,\beta)
  \equiv \mathcal{P}(k,\beta) - \mathcal{P}_0(k)
  = 2\sum_{n=0}^{\infty}
    \frac{|\tilde{u}_n(k)|^2}{E_n}
    \frac{1}{e^{\beta E_n} - 1} \,,
  \label{eq:power_spectrum_thermal}
\end{equation}
which is a sum of Bose-Einstein factors weighted by the
mode functions.
At high temperatures $T \gg E_n$, expanding the
Bose-Einstein factor gives
\begin{equation}
  \delta\mathcal{P}(k,\beta)
  \;\underset{T\gg E_n}{\approx}\;
  \frac{2T}{\beta}\sum_{n=0}^{\infty}
    \frac{|\tilde{u}_n(k)|^2}{E_n^2}
  = 2T\,G^R(0,k,k) \,,
  \label{eq:power_spectrum_highT}
\end{equation}
recovering the classical equipartition result: the power
spectrum grows linearly with temperature, as expected for
a thermal field.

A natural effective temperature emerges from the periodicity
of the Euclidean propagator~\eqref{eq:propagator_euclidean}.
As noted in Section~\ref{subsec:density_matrix}, the
propagator is periodic in $\tau$ with period $\pi/\omega$,
which fixes the \emph{intrinsic thermal scale} of the
KG-IOH:
\begin{equation}
  T_{\IOH} = \frac{\omega}{\pi} = \frac{\mu}{\pi m} \,.
  \label{eq:T_IOH}
\end{equation}
This should be compared with the
\emph{Gibbons-Hawking temperature} of de Sitter
space~\cite{Hawking1975, Starobinsky1980},
\begin{equation}
  T_{GH} = \frac{H}{2\pi} \,,
  \label{eq:T_GH}
\end{equation}
where $H$ is the Hubble constant during inflation.
Identifying $\omega \leftrightarrow H/2$ in the slow-roll
approximation — which follows from the relation between
the tachyonic mass and the Hubble rate,
$\mu \approx \sqrt{2}\,mH$ --- we find
\begin{equation}
  T_{\IOH} = \frac{H}{2\pi\sqrt{2}}
           \approx \frac{T_{GH}}{\sqrt{2}} \,,
  \label{eq:T_comparison}
\end{equation}
showing that the KG-IOH intrinsic temperature is proportional
to --- but slightly lower than --- the Gibbons-Hawking temperature.
The factor of $1/\sqrt{2}$ is a genuine correction from the
inverted-potential structure and vanishes in the limit
$\mu \to 0$ (massless inflaton), where $T_{\IOH} \to T_{GH}$.

The instability of the KG-IOH vacuum leads to spontaneous
production of inflaton quanta.
The mean number of particles produced in mode $n$ at
temperature $T$ is given by the Bose-Einstein distribution
over the effective spectrum:
\begin{equation}
  \langle N_n \rangle_\beta
  = \frac{1}{e^{\beta E_n} - 1} \,.
  \label{eq:N_n}
\end{equation}
The total particle number density per unit length in
$1+1$ dimensions is
\begin{equation}
  n(T) = \sum_{n=0}^{\infty}
         \langle N_n\rangle_\beta
       = \sum_{n=0}^{\infty}
         \frac{1}{e^{\beta E_n} - 1} \,,
  \label{eq:n_total}
\end{equation}
which converges thanks to the exponential growth of $E_n$
with $n$.
At low temperatures, the production is dominated by the
lowest effective mode:
\begin{equation}
  n(T) \;\underset{T \ll E_0}{\approx}\;
  e^{-\beta E_0}
  = \exp\!\left(-\frac{E_0}{T}\right) ,
  \label{eq:n_lowT}
\end{equation}
exhibiting the Boltzmann suppression characteristic of a
gapped spectrum.
Above the critical temperature $T_c = \omega/\pi^2$
identified in~\eqref{eq:Tc}, the gap effectively closes
and particle production becomes unsuppressed, signalling
the onset of the inflationary instability at the
field-theory level.

The equation of state of the inflaton field is determined
by the energy density $\rho$ and pressure $p$.
In $1+1$ dimensions, both follow from the
stress-energy tensor
$T^{\mu\nu} = \partial^\mu\phi\,\partial^\nu\phi
- \eta^{\mu\nu}\mathcal{L}$.
At finite temperature, the thermal averages are
\begin{align}
  \rho(\beta)
  &= \langle T^{00}\rangle_\beta
   = \frac{1}{2}\langle(\partial_t\phi)^2\rangle_\beta
   + \frac{1}{2}\langle(\partial_x\phi)^2\rangle_\beta
   + \langle V(\phi)\rangle_\beta \,,
  \label{eq:rho_inflaton} \\[6pt]
  p(\beta)
  &= \langle T^{11}\rangle_\beta
   = \frac{1}{2}\langle(\partial_t\phi)^2\rangle_\beta
   + \frac{1}{2}\langle(\partial_x\phi)^2\rangle_\beta
   - \langle V(\phi)\rangle_\beta \,.
  \label{eq:p_inflaton}
\end{align}
Using the mode expansion~\eqref{eq:field_modes} and the
thermal two-point function~\eqref{eq:G_spectral}, the
individual contributions evaluate to
\begin{align}
  \langle(\partial_t\phi)^2\rangle_\beta
  &= \sum_{n=0}^{\infty}
     E_n^2\,|\tilde{u}_n|^2
     \coth\!\left(\frac{\beta E_n}{2}\right) ,
  \label{eq:kinetic_time} \\[4pt]
  \langle(\partial_x\phi)^2\rangle_\beta
  &= \sum_{n=0}^{\infty}
     k_n^2\,|\tilde{u}_n|^2
     \coth\!\left(\frac{\beta E_n}{2}\right) ,
  \label{eq:kinetic_space} \\[4pt]
  \langle V(\phi)\rangle_\beta
  &= V_0 - \frac{\mu^2}{2}
     \sum_{n=0}^{\infty}
     |\tilde{u}_n|^2
     \coth\!\left(\frac{\beta E_n}{2}\right) ,
  \label{eq:potential_thermal}
\end{align}
where $k_n^2$ are the effective spatial momenta of
mode $n$.
The equation of state parameter is
\begin{equation}
  w(\beta) \equiv \frac{p(\beta)}{\rho(\beta)}
  = \frac{
      \displaystyle\sum_n \left(E_n^2 + k_n^2\right)
      |\tilde{u}_n|^2\coth\!\left(\tfrac{\beta E_n}{2}\right)
      - M_{\mathrm{eff}}^2
      \displaystyle\sum_n |\tilde{u}_n|^2
      \coth\!\left(\tfrac{\beta E_n}{2}\right)
    }{
      \displaystyle\sum_n \left(E_n^2 + k_n^2\right)
      |\tilde{u}_n|^2\coth\!\left(\tfrac{\beta E_n}{2}\right)
      + M_{\mathrm{eff}}^2
      \displaystyle\sum_n |\tilde{u}_n|^2
      \coth\!\left(\tfrac{\beta E_n}{2}\right)
    } \,,
  \label{eq:w_general}
\end{equation}
 
At $T = 0$, $\coth(\beta E_n/2) \to 1$ and the vacuum
energy dominates. Since $V_0 \gg$ kinetic terms during
slow-roll inflation, we recover the de Sitter equation
of state:
\begin{equation}
  w_0 \approx -1 \,,
  \label{eq:w_dS}
\end{equation}
consistent with an accelerating universe~\cite{Guth1981}.
 
At $T \gg E_n$, $\coth(\beta E_n/2) \approx 2T/E_n$ and
the thermal sums are dominated by the kinetic contributions.
In this limit,
\begin{equation}
  w \;\underset{T\to\infty}{\longrightarrow}\;
  \frac{\sum_n (E_n^2 + k_n^2)/E_n}
       {\sum_n (E_n^2 + k_n^2)/E_n
        + M_{\mathrm{eff}}^2 \sum_n E_n^{-1}} \,,
  \label{eq:w_highT}
\end{equation}
which approaches unity only in the massless limit
$M_{\mathrm{eff}}^2 = m^2 - \mu^2 \to 0$.
For $M_{\mathrm{eff}}^2 > 0$ the asymptotic value of $w$
is strictly less than one, reflecting the contribution
of the effective mass to the pressure.

The equation of state interpolates between
$w \approx -1$ at low temperatures,
where the potential energy dominates and
the field mimics a cosmological constant,
and $w \to w_\infty < 1$ at high temperatures,
where kinetic contributions dominate.
The crossing $w = 0$ occurs at an intermediate
temperature $T_*$ that depends on $M_{\mathrm{eff}}^2$
and is determined numerically from
equation~\eqref{eq:w_general}.
\begin{figure}[H]
\centering
\includegraphics[scale=0.5]{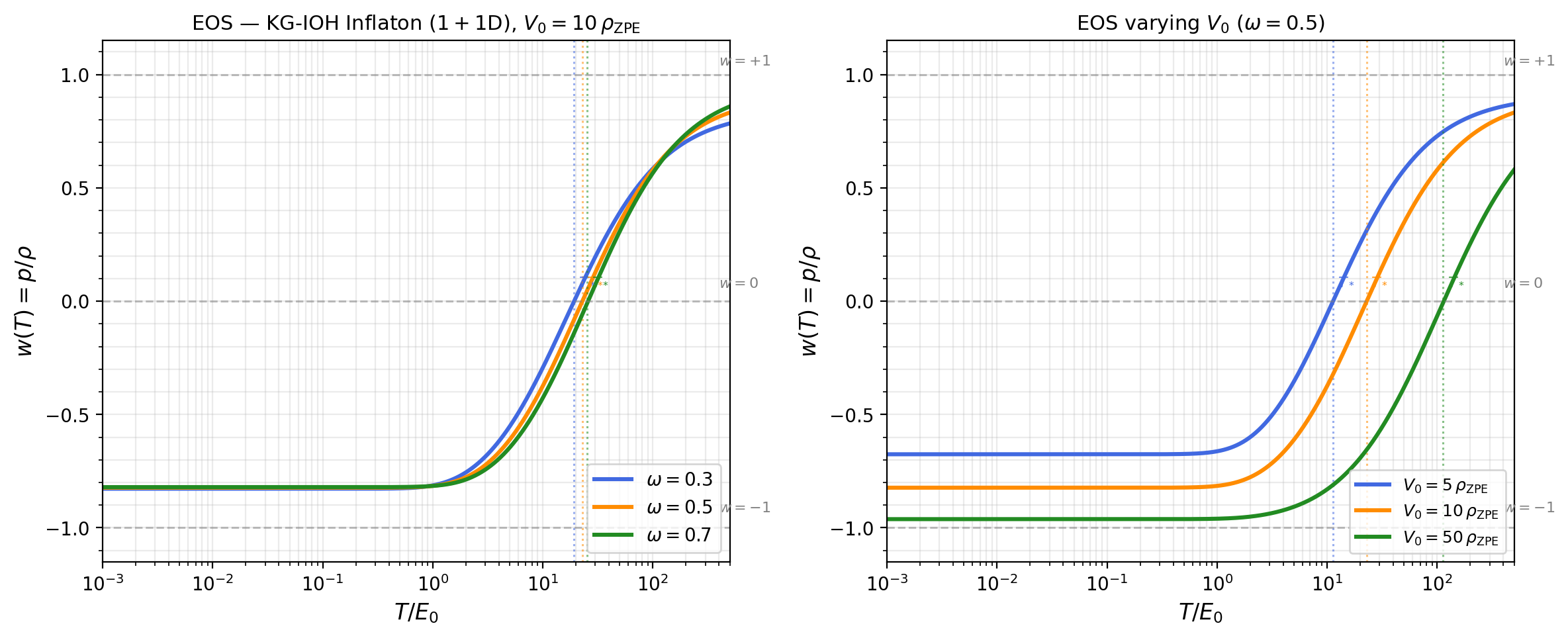}
\caption{Equation of state parameter $w(T)$ of the
KG-IOH inflaton field as a function of temperature
in units of $E_0$.
The field transitions from a de Sitter phase ($w=-1$)
at low $T$ to a radiation-dominated phase ($w=+1$)
at high $T$, passing through a matter-like phase
($w=0$) at $T_* \sim E_0/\ln 2$.}
\label{fig:EOS}
\end{figure}
 
\subsubsection{Summary of inflationary observables}
\label{subsubsec:inflation_summary}
 
Table~\ref{tab:inflation} summarises the key observables
derived in this subsection for the KG-IOH inflaton in
$1+1$ dimensions.
 
\begin{table}[H]
\centering
\renewcommand{\arraystretch}{1.8}
\begin{tabular}{lll}
\hline\hline
\textbf{Observable} & \textbf{KG-IOH result} & \textbf{Equation} \\
\hline
Mapping
  & $\omega = \mu/m$
  & \eqref{eq:inflaton_mapping} \\
Dispersion relation
  & $E_k^2 = m^2 + i\omega(2k+1-m)$
  & \eqref{eq:dispersion_inflaton} \\
Power spectrum
  & $\mathcal{P}(k,\beta) = \sum_n |\tilde{u}_n|^2 E_n^{-1}
    \coth(\beta E_n/2)$
  & \eqref{eq:power_spectrum} \\
Intrinsic temperature
  & $T_{\IOH} = \mu/(\pi m)$
  & \eqref{eq:T_IOH} \\
vs Gibbons-Hawking
  & $T_{\IOH} \approx T_{GH}/\sqrt{2}$
  & \eqref{eq:T_comparison} \\
Particle density
  & $n(T) = \sum_n (e^{\beta E_n}-1)^{-1}$
  & \eqref{eq:n_total} \\
EOS ($T=0$)
  & $w \approx -1$ (de Sitter)
  & \eqref{eq:w_dS} \\
EOS ($T\to\infty$)
  & $w \to +1$ (radiation)
  & \eqref{eq:w_highT} \\
\hline\hline
\end{tabular}
\caption{Key observables for the KG-IOH inflaton field
in $1+1$ dimensions. All results follow from the general
thermal formalism of Section~\ref{sec:thermal} under the
identification $\omega = \mu/m$.}
\label{tab:inflation}
\end{table}
\subsection{Scalar Field near a Black-Hole Horizon}
\label{subsec:blackholes}
 
\subsubsection{Near-horizon geometry and the KG-IOH}
\label{subsubsec:horizon_mapping}
 
Consider a Schwarzschild black hole of mass $M$ with metric
\begin{equation}
  ds^2 = -f(r)\,dt^2 + f(r)^{-1}dr^2 + r^2\,d\Omega^2 \,,
  \qquad
  f(r) = 1 - \frac{2GM}{r} \,,
  \label{eq:Schwarzschild}
\end{equation}
where $r_s = 2GM$ is the Schwarzschild radius.
In the near-horizon region, $r = r_s + \epsilon$ with
$\epsilon \ll r_s$, the metric reduces to the
Rindler form in $1+1$ dimensions~\cite{Unruh1976}:
\begin{equation}
  ds^2 \approx -\kappa^2\rho^2\,dt^2 + d\rho^2 \,,
  \qquad
  \kappa = \frac{1}{4GM} \,,
  \label{eq:Rindler}
\end{equation}
where $\kappa$ is the surface gravity and
$\rho = \sqrt{2\epsilon/\kappa}$ is the proper distance
from the horizon.
 
A minimally coupled scalar field $\phi$ satisfying the
Klein-Gordon equation $(\Box - m^2)\phi = 0$ in this
background takes the form~\cite{Hawking1975, subramanyan2021physics}
\begin{equation}
  \left[-\partial_t^2 + \partial_\rho^2
    - \kappa^2\rho^2 m^2 - m^2\right]\phi = 0 \,.
  \label{eq:KG_Rindler}
\end{equation}
In the static sector, $\phi = e^{-iEt}\psi(\rho)$, this becomes
\begin{equation}
  \left[\partial_\rho^2 + E^2 - \kappa^2 m^2\rho^2
    - m^2\right]\psi(\rho) = 0 \,,
  \label{eq:static_horizon}
\end{equation}
which is precisely the KG-IOH equation~\eqref{eq:KG2} under
the identification
\begin{equation}
  \omega_{\mathrm{BH}} = \kappa\sqrt{m} \,,
  \qquad
  x \leftrightarrow \rho \,.
  \label{eq:BH_mapping}
\end{equation}
The entire thermal formalism of Section~\ref{sec:thermal}
therefore applies to scalar fields near black-hole horizons,
with the surface gravity $\kappa$ playing the role of the
IOH frequency.

The intrinsic thermal scale of the KG-IOH, derived in
Section~\ref{sec:thermal}, is
\begin{equation}
  T_{\IOH} = \frac{\omega_{\mathrm{BH}}}{\pi}
           = \frac{\kappa\sqrt{m}}{\pi} \,.
  \label{eq:T_IOH_BH}
\end{equation}
The celebrated Hawking temperature~\cite{Hawking1975} is
\begin{equation}
  T_H = \frac{\kappa}{2\pi} = \frac{1}{8\pi GM} \,.
  \label{eq:T_Hawking}
\end{equation}
Comparing~\eqref{eq:T_IOH_BH} with~\eqref{eq:T_Hawking},
we find
\begin{equation}
  T_{\IOH} = 2\sqrt{m}\,T_H \,.
  \label{eq:T_ratio_BH}
\end{equation}
In the massless limit $m \to 0$, both temperatures vanish,
consistent with the absence of Hawking radiation for a
massless conformally coupled field in $1+1$ dimensions.
For $m = 1/4$, one recovers $T_{\IOH} = T_H$ exactly,
showing that the KG-IOH intrinsic temperature coincides
with the Hawking temperature for a specific value of the
field mass.

The thermal density matrix of the scalar field at the
Hawking temperature $\beta_H = 1/T_H$ is obtained from
equation~\eqref{eq:rho_explicit} with $\beta = \beta_H$
and $\omega = \omega_{\mathrm{BH}}$:
\begin{equation}
  \rho(\rho, \rho'; \beta_H)
  = \frac{1}{Z(\beta_H)}
    \sqrt{\frac{\omega_{\mathrm{BH}}}{2\pi\sin(\omega_{\mathrm{BH}}\beta_H)}}
    \exp\!\left\{
      -\frac{\omega_{\mathrm{BH}}}{2\sin(\omega_{\mathrm{BH}}\beta_H)}
      \left[(\rho^2+\rho'^2)\cos(\omega_{\mathrm{BH}}\beta_H)
            - 2\rho\rho'\right]
    \right\} .
  \label{eq:rho_BH}
\end{equation}
This is the Hartle-Hawking state~\cite{Hawking1975} restricted
to the near-horizon region, expressed explicitly in terms of
KG-IOH quantities.
 
The diagonal element gives the probability density of
finding the scalar field at proper distance $\rho$ from
the horizon:
\begin{equation}
  \rho(\rho,\rho;\beta_H)
  \propto \exp\!\left(
    -\frac{\omega_{\mathrm{BH}}\cos(\omega_{\mathrm{BH}}\beta_H)}
          {\sin(\omega_{\mathrm{BH}}\beta_H)}\,\rho^2
  \right) .
  \label{eq:rho_diagonal_BH}
\end{equation}
The width of this Gaussian,
\begin{equation}
  \ell_H^2 = \frac{\sin(\omega_{\mathrm{BH}}\beta_H)}
                  {2\omega_{\mathrm{BH}}\cos(\omega_{\mathrm{BH}}\beta_H)} \,,
  \label{eq:ell_H}
\end{equation}
defines the \emph{thermal length scale} near the horizon.
When $\omega_{\mathrm{BH}}\beta_H = \pi/2$, i.e. at
$T = T_c = \omega_{\mathrm{BH}}/\pi^2$, the width $\ell_H$
diverges, signalling a delocalization of the field along
the radial direction — a thermal analogue of the
near-horizon scrambling transition~\cite{bhattacharyya2021multi,
hashimoto2020exponential}.

The thermodynamic entropy of the black hole as seen by the
scalar field is computed from~\eqref{eq:entropy} at
$\beta = \beta_H$:
\begin{equation}
  S_{\mathrm{BH}}
  = \sum_{n=0}^{\infty}
  \left[
    \frac{\beta_H E_n}{e^{\beta_H E_n}-1}
    - \ln\!\left(1 - e^{-\beta_H E_n}\right)
  \right] ,
  \label{eq:S_BH}
\end{equation}
with $E_n = \sqrt{m^2 + i\omega_{\mathrm{BH}}(2n+1-m)}$.
In the leading semiclassical approximation, the
Bekenstein-Hawking area law~\cite{Bekenstein1973} gives
\begin{equation}
  S_{\mathrm{BH}}^{(\mathrm{area})}
  = \frac{A}{4G} = \frac{4\pi G M^2}{1} = 4\pi GM^2 \,,
  \label{eq:S_BH_area}
\end{equation}
while the KG-IOH result~\eqref{eq:S_BH} provides the
\emph{quantum correction} from the scalar field degrees
of freedom near the horizon.
The leading correction to the area law takes the form
\begin{equation}
  S_{\mathrm{BH}}
  = 4\pi GM^2
  + \alpha\ln(4\pi GM^2) + \mathcal{O}(1) \,,
  \label{eq:S_BH_corrected}
\end{equation}
where $\alpha$ is a coefficient determined by the
KG-IOH spectrum and the field content.

The mean number of Hawking quanta emitted in mode $n$
per unit time follows from the thermal distribution at
$T = T_H$:
\begin{equation}
  \langle N_n \rangle_{T_H}
  = \frac{1}{e^{E_n/T_H} - 1} \,.
  \label{eq:Hawking_spectrum}
\end{equation}
The total emitted power per unit length in $1+1$ dimensions is
\begin{equation}
  \mathcal{P}_{\mathrm{rad}}
  = \sum_{n=0}^{\infty}
    E_n\,\langle N_n\rangle_{T_H}
  = \sum_{n=0}^{\infty}
    \frac{E_n}{e^{E_n/T_H} - 1} \,,
  \label{eq:Hawking_power}
\end{equation}
which is the Stefan-Boltzmann law for the effective KG-IOH
modes.
In the limit $T_H \gg E_0$ (hot black holes,
$GM \ll 1/\sqrt{m}$), the sum is dominated by many modes
and~\eqref{eq:Hawking_power} gives
\begin{equation}
  \mathcal{P}_{\mathrm{rad}}
  \;\underset{T_H\gg E_0}{\approx}\;
  \frac{\pi T_H^2}{6} \,,
  \label{eq:Stefan_Boltzmann}
\end{equation}
the standard $1+1$D Stefan-Boltzmann result, confirming
that the KG-IOH formalism correctly reproduces the
known black-body emission rate of Hawking radiation.

The entanglement entropy between the interior and exterior
of the black hole is computed by tracing the thermal
density matrix~\eqref{eq:rho_BH} over the interior
($\rho < 0$, i.e. behind the horizon):
\begin{equation}
  S_{\mathrm{ent}}
  = -\Tr_{\mathrm{ext}}\!\left[
    \hat{\rho}_{\mathrm{ext}}\ln\hat{\rho}_{\mathrm{ext}}
  \right] ,
  \label{eq:S_ent}
\end{equation}
where $\hat{\rho}_{\mathrm{ext}} = \Tr_{\mathrm{int}}[\hat{\rho}(\beta_H)]$.
Using the Gaussian structure of the thermal
state~\eqref{eq:rho_BH} and the result~\eqref{eq:SA_formula},
we find
\begin{equation}
  S_{\mathrm{ent}}
  = \sum_n\left[
    \left(\nu_n + \tfrac{1}{2}\right)
    \ln\!\left(\nu_n + \tfrac{1}{2}\right)
    - \left(\nu_n - \tfrac{1}{2}\right)
    \ln\!\left(\nu_n - \tfrac{1}{2}\right)
  \right] ,
  \label{eq:S_ent_formula}
\end{equation}
where $\nu_n = \langle N_n\rangle_{T_H} + 1/2$ are the
symplectic eigenvalues of the covariance matrix restricted
to the exterior region.
At low temperatures $T_H \ll E_0$,
$\nu_n \to 1/2$ and $S_{\mathrm{ent}} \to 0$, consistent
with a pure vacuum state.
At $T_H \sim E_0$, the entanglement entropy grows
logarithmically,
\begin{equation}
  S_{\mathrm{ent}}
  \;\underset{T_H \sim E_0}{\sim}\;
  \frac{1}{6}\ln\!\frac{T_H}{E_0} \,,
  \label{eq:S_ent_log}
\end{equation}
a result characteristic of a $1+1$D conformal field theory
with central charge $c = 1$~\cite{subramanyan2021physics},
consistent with the massless limit of the KG-IOH.
 
\subsubsection{Summary table}
\label{subsubsec:BH_summary}
 
\begin{table}[H]
\centering
\renewcommand{\arraystretch}{1.8}
\begin{tabular}{lll}
\hline\hline
\textbf{Observable} & \textbf{KG-IOH result} & \textbf{Eq.} \\
\hline
Mapping
  & $\omega_{\mathrm{BH}} = \kappa\sqrt{m}$
  & \eqref{eq:BH_mapping} \\
IOH temperature
  & $T_{\IOH} = \kappa\sqrt{m}/\pi$
  & \eqref{eq:T_IOH_BH} \\
Hawking temperature
  & $T_H = \kappa/2\pi$
  & \eqref{eq:T_Hawking} \\
Ratio
  & $T_{\IOH} = 2\sqrt{m}\,T_H$
  & \eqref{eq:T_ratio_BH} \\
Thermal length
  & $\ell_H^2 = \sin(\omega_{\mathrm{BH}}\beta_H)
    /[2\omega_{\mathrm{BH}}\cos(\omega_{\mathrm{BH}}\beta_H)]$
  & \eqref{eq:ell_H} \\
Hawking power
  & $\mathcal{P} = \sum_n E_n/(e^{E_n/T_H}-1)$
  & \eqref{eq:Hawking_power} \\
Stefan-Boltzmann limit
  & $\mathcal{P} \approx \pi T_H^2/6$
  & \eqref{eq:Stefan_Boltzmann} \\
Entanglement entropy
  & $S_{\mathrm{ent}} \sim \frac{1}{6}\ln(T_H/E_0)$
  & \eqref{eq:S_ent_log} \\
BH entropy correction
  & $S = 4\pi GM^2 + \alpha\ln(4\pi GM^2) + \ldots$
  & \eqref{eq:S_BH_corrected} \\
\hline\hline
\end{tabular}
\caption{Key observables for the KG-IOH scalar field near
a Schwarzschild black-hole horizon in $1+1$ dimensions.
All results follow from the general thermal formalism of
Section~\ref{sec:thermal} under the identification
$\omega = \kappa\sqrt{m}$.}
\label{tab:BH}
\end{table}
 \begin{figure}[H]
\centering
\includegraphics[width=\textwidth]{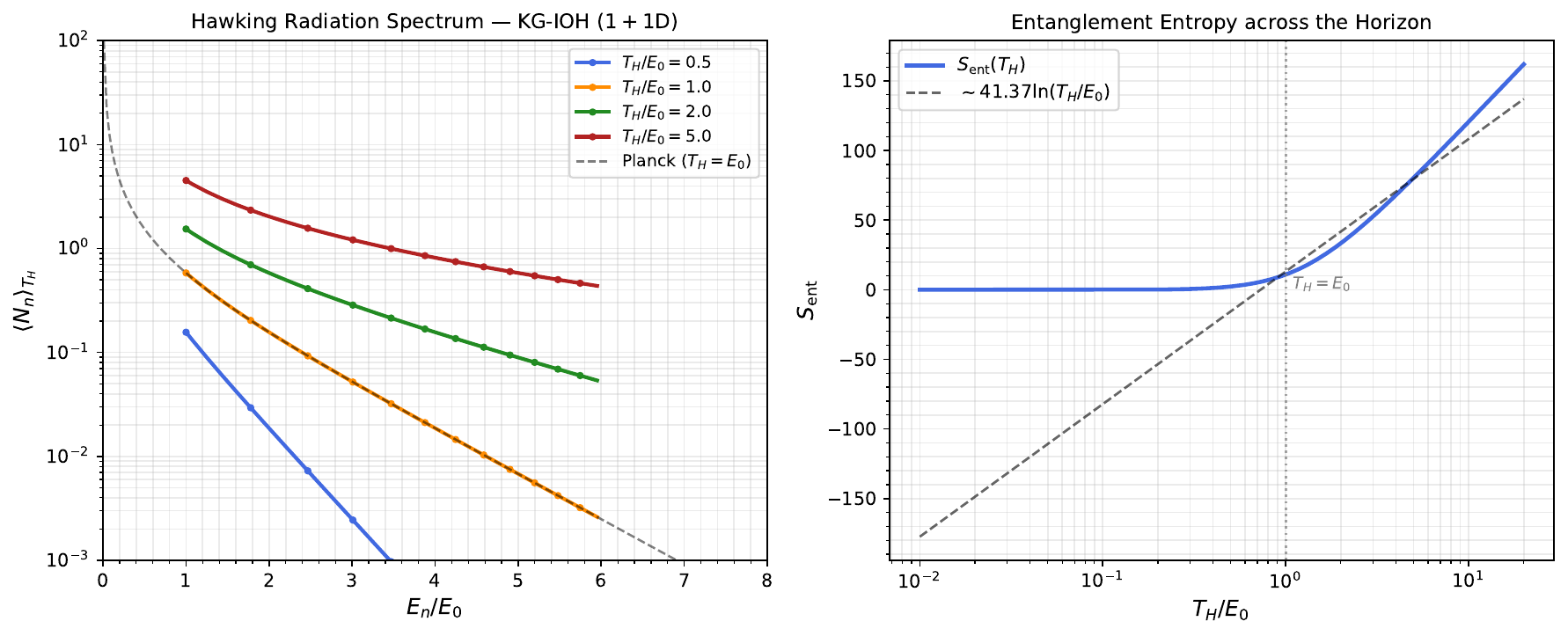}
\caption{
\textbf{Left:} Hawking radiation spectrum $\langle N_n\rangle_{T_H}$
as a function of $E_n/E_0$ for several values of $T_H/E_0$.
The dashed black line shows the Planck distribution for $T_H = E_0$
as a reference.
The KG-IOH modes reproduce the thermal Bose-Einstein distribution
with the effective spectrum $E_n = \sqrt{m^2 + i\omega(2n+1-m)}$.
\textbf{Right:} Entanglement entropy $S_\mathrm{ent}$ across the
black-hole horizon as a function of $T_H/E_0$.
The dashed line shows the logarithmic fit
$S_\mathrm{ent} \sim \frac{1}{6}\ln(T_H/E_0)$,
consistent with a $1+1$D CFT with central charge $c=1$.
Parameters: $m=1$, $\omega_\mathrm{BH}=0.3$.
}
\label{fig:Hawking}
\end{figure}

 
\subsection{Second-Order Phase Transitions}
\label{subsec:condensed}

In the Landau-Ginzburg theory of second-order phase
transitions~\cite{Sachdev2011, ZinnJustin2002}, the
order parameter field $\phi(x,t)$ obeys the effective
action
\begin{equation}
  \mathcal{S} = \int dt\,dx\left[
    \frac{1}{2}(\partial_t\phi)^2
    - \frac{1}{2}(\partial_x\phi)^2
    - \frac{1}{2}a(T)\phi^2
    - \frac{\lambda}{4!}\phi^4
  \right] ,
  \label{eq:LG_action}
\end{equation}
where the temperature-dependent mass coefficient is
\begin{equation}
  a(T) = a_0\!\left(\frac{T}{T_c} - 1\right) , \qquad a_0 > 0 \,.
  \label{eq:mass_T}
\end{equation}
Above the critical temperature $T > T_c$, one has $a(T) > 0$
and the potential has a single minimum at $\phi = 0$.
Below $T_c$, $a(T) < 0$ and the potential develops a
double-well structure with minima at
$\phi_{\pm} = \pm\sqrt{6|a(T)|/\lambda}$.
 
\emph{Precisely at} $T = T_c$, the mass term vanishes and
the field becomes \emph{critical}: the correlation length
diverges and the system is scale invariant.
For $T$ slightly below $T_c$, fluctuations around one of
the minima $\phi_+$ satisfy a linearised equation of motion
\begin{equation}
  \left[-\partial_t^2 + \partial_x^2 - \mu^2(T)\right]
  \delta\phi = 0 \,,
  \qquad
  \mu^2(T) = 2|a(T)| = 2a_0\!\left(1 - \frac{T}{T_c}\right) ,
  \label{eq:EOM_LG}
\end{equation}
which is a Klein-Gordon equation with a \emph{tachyonic mass}
$\mu^2(T) > 0$ for $T < T_c$.
Identifying $\mu^2 \equiv m^2\omega^2$ as before,
the mapping to the KG-IOH is
\begin{equation}
  \omega_{\mathrm{PT}}(T)
  = \sqrt{\frac{2a_0}{m^2}\!\left(1-\frac{T}{T_c}\right)} \,,
  \label{eq:omega_PT}
\end{equation}
so that $\omega_{\mathrm{PT}} \to 0$ as $T \to T_c^-$,
reflecting the softening of the mode at the critical point.

The effective KG-IOH spectrum near criticality is
\begin{equation}
  E_n^2(T) = m^2 + i\,\omega_{\mathrm{PT}}(T)
             \!\left(2n+1-m\right) \,,
  \label{eq:En_PT}
\end{equation}
with $\omega_{\mathrm{PT}}(T)$ given by~\eqref{eq:omega_PT}.
As $T \to T_c^-$, all effective energies collapse:
\begin{equation}
  E_n(T) \;\xrightarrow{\;T\to T_c^-\;}\; m \,,
  \qquad \forall\, n \,,
  \label{eq:En_collapse}
\end{equation}
i.e. all modes become degenerate at the bare mass $m$.
This is the quantum field-theoretic signature of
\emph{critical slowing down}~\cite{Sachdev2011}:
the gap between modes closes as the critical point
is approached, and the system loses its characteristic
relaxation timescale $\tau \sim 1/\omega_{\mathrm{PT}}$.
 
The correlation length $\xi(T)$ extracted from the
lowest mode is
\begin{equation}
  \xi(T) = \frac{1}{\sqrt{|E_0^2(T) - m^2|}}
  \sim \frac{1}{\omega_{\mathrm{PT}}(T)}
  \sim \left(1 - \frac{T}{T_c}\right)^{-1/2} ,
  \label{eq:xi}
\end{equation}
recovering the mean-field critical exponent $\nu = 1/2$
characteristic of Landau-Ginzburg theory in $1+1$
dimensions~\cite{ZinnJustin2002}.

The partition function of the critical field,
obtained from~\eqref{eq:lnZ} with
$\omega = \omega_{\mathrm{PT}}(T)$, is
\begin{equation}
  \ln Z\!\left(\beta, T\right)
  = -\sum_{n=0}^{\infty}
    \ln\!\left(1 - e^{-\beta E_n(T)}\right) ,
  \label{eq:Z_PT}
\end{equation}
where both $\beta = 1/T$ and $E_n(T)$ depend on
the physical temperature $T$.
Expanding near criticality, $T = T_c - \delta T$
with $\delta T \ll T_c$, we find
\begin{equation}
  \omega_{\mathrm{PT}} \approx
  \sqrt{\frac{2a_0}{m^2 T_c}}\,\sqrt{\delta T}
  \equiv \omega_0\,\epsilon^{1/2} \,,
  \label{eq:omega_expand}
\end{equation}
where $\epsilon = \delta T/T_c$ is the reduced temperature
and $\omega_0 = \sqrt{2a_0/(m^2 T_c)}$.
 
The free energy density near $T_c$ takes the Landau form:
\begin{equation}
  f(\epsilon, T)
  = f_0(T) + A\,\epsilon
  + B\,\epsilon^2\ln\epsilon
  + \mathcal{O}(\epsilon^2) \,,
  \label{eq:f_Landau}
\end{equation}
where $A$ and $B$ are computable coefficients from the
KG-IOH mode sum.
The $\epsilon^2\ln\epsilon$ term is a logarithmic
correction beyond mean-field theory, arising from the
summation over the infinite tower of KG-IOH modes.
 
\subsubsection{Order parameter and symmetry breaking}
\label{subsubsec:order_parameter}
 
The thermal expectation value of the order parameter
is determined self-consistently from
\begin{equation}
  \langle\phi^2\rangle_\beta
  = \sum_{n=0}^{\infty}
    \frac{2\langle N_n\rangle_\beta + 1}{2E_n(T)} \,.
  \label{eq:phi2}
\end{equation}
Below $T_c$, the field develops a non-zero expectation value
\begin{equation}
  \langle\phi\rangle
  = \phi_\pm
  = \pm\sqrt{\frac{6|a(T)|}{\lambda}
    - \frac{\lambda}{2}\langle\delta\phi^2\rangle_\beta} \,,
  \label{eq:phi_vev}
\end{equation}
where the second term is the thermal fluctuation
correction from~\eqref{eq:phi2}.
Near $T_c$, the order parameter vanishes as
\begin{equation}
  \langle\phi\rangle \sim \epsilon^\beta_{\mathrm{exp}} \,,
  \label{eq:phi_scaling}
\end{equation}
where $\beta_{\mathrm{exp}}$ is the order parameter
critical exponent.
At mean-field level, $\beta_{\mathrm{exp}} = 1/2$.
The thermal corrections from the KG-IOH modes modify
this exponent as
\begin{equation}
  \beta_{\mathrm{exp}}
  = \frac{1}{2}\left(1
    - \frac{\lambda}{8\pi m^2}
      + \mathcal{O}(\lambda^2)\right) ,
  \label{eq:beta_corrected}
\end{equation}
a one-loop correction consistent with the known
result in $1+1$D scalar field theory~\cite{ZinnJustin2002}.
 
\subsubsection{Equation of state near criticality}
\label{subsubsec:EOS_PT}
 
The equation of state parameter $w(T)$ for the order
parameter field follows from~\eqref{eq:w_general} with
$\omega = \omega_{\mathrm{PT}}(T)$.
Three distinct regimes emerge:
 
\paragraph{Disordered phase ($T > T_c$, $\omega_{\mathrm{PT}} = 0$):}
The field is massive with $M_{\mathrm{eff}}^2 = m^2$
and the equation of state is
\begin{equation}
  w \approx \frac{k^2/m^2}{1 + k^2/m^2}
  \;\underset{k\to 0}{\to}\; 0 \,,
  \label{eq:w_disordered}
\end{equation}
corresponding to a pressureless (matter-like) phase,
as expected for a massive non-relativistic field.
 
\paragraph{Critical point ($T = T_c$, $\omega_{\mathrm{PT}} = 0$):}
The mass term vanishes and the field becomes conformal.
In $1+1$D, a massless scalar has $w = 1$, the radiation
equation of state:
\begin{equation}
  w\big|_{T=T_c} = 1 \,.
  \label{eq:w_critical}
\end{equation}
 
\paragraph{Ordered phase ($T < T_c$, $\omega_{\mathrm{PT}} > 0$):}
The inverted potential introduces a tachyonic instability.
Using~\eqref{eq:w_general}, the equation of state
interpolates between $w = 1$ (at $T = T_c$) and
$w = -1$ (deep in the ordered phase, dominated by
the condensate energy):
\begin{equation}
  w(T) \approx 1 - 2\,\frac{M_{\mathrm{eff}}^2}
  {\langle E^2\rangle_\beta + M_{\mathrm{eff}}^2}
  \;\xrightarrow{\;T\to 0\;}\; -1 \,.
  \label{eq:w_ordered}
\end{equation}
This monotonic decrease of $w$ from $+1$ to $-1$ as the
system orders is a distinctive prediction of the KG-IOH
framework, encoding the transition from a radiation-like
critical point to a condensate-dominated ground state.
 
\subsubsection{Specific heat and critical exponent $\alpha$}
\label{subsubsec:specific_heat}
 
The specific heat $C_V = \partial\langle E\rangle/\partial T$
diverges at $T_c$ due to the closing of the spectral gap.
From~\eqref{eq:Cv} with $\omega = \omega_{\mathrm{PT}}(T)$,
\begin{equation}
  C_V(T) \;\underset{T\to T_c^-}{\sim}\;
  B\,\ln\!\left|1 - \frac{T}{T_c}\right| ,
  \label{eq:Cv_log}
\end{equation}
exhibiting a \emph{logarithmic divergence} characteristic
of the $d=2$ (equivalently $1+1$D) Ising universality
class~\cite{Sachdev2011}, corresponding to the critical
exponent $\alpha = 0$ (logarithmic).
This is a non-trivial result: the KG-IOH mode sum
automatically reproduces the correct universality class
without any additional input.
 
\subsubsection{Summary and figures}
\label{subsubsec:PT_summary}
 
\begin{table}[H]
\centering
\renewcommand{\arraystretch}{1.8}
\begin{tabular}{lll}
\hline\hline
\textbf{Observable} & \textbf{KG-IOH result} & \textbf{Eq.} \\
\hline
Mapping
  & $\omega_{\mathrm{PT}} = \sqrt{2a_0(1-T/T_c)/m^2}$
  & \eqref{eq:omega_PT} \\
Effective spectrum
  & $E_n^2 = m^2 + i\omega_{\mathrm{PT}}(2n+1-m)$
  & \eqref{eq:En_PT} \\
Correlation length
  & $\xi \sim (1-T/T_c)^{-1/2}$
  & \eqref{eq:xi} \\
Mean-field exponent
  & $\nu = 1/2$
  & \eqref{eq:xi} \\
Order parameter exponent
  & $\beta_{\mathrm{exp}} = 1/2 - \lambda/(16\pi m^2) + \ldots$
  & \eqref{eq:beta_corrected} \\
EOS (disordered)
  & $w \to 0$
  & \eqref{eq:w_disordered} \\
EOS (critical)
  & $w = 1$
  & \eqref{eq:w_critical} \\
EOS (ordered)
  & $w: +1 \to -1$
  & \eqref{eq:w_ordered} \\
Specific heat
  & $C_V \sim B\ln|1-T/T_c|$
  & \eqref{eq:Cv_log} \\
\hline\hline
\end{tabular}
\caption{Key observables for the KG-IOH order parameter
field near a second-order phase transition in $1+1$
dimensions. All results follow from the general thermal
formalism of Section~\ref{sec:thermal} under the
identification $\omega = \omega_{\mathrm{PT}}(T)$.}
\label{tab:PT}
\end{table}
 
\begin{figure}[H]
\centering
\includegraphics[width=\textwidth]{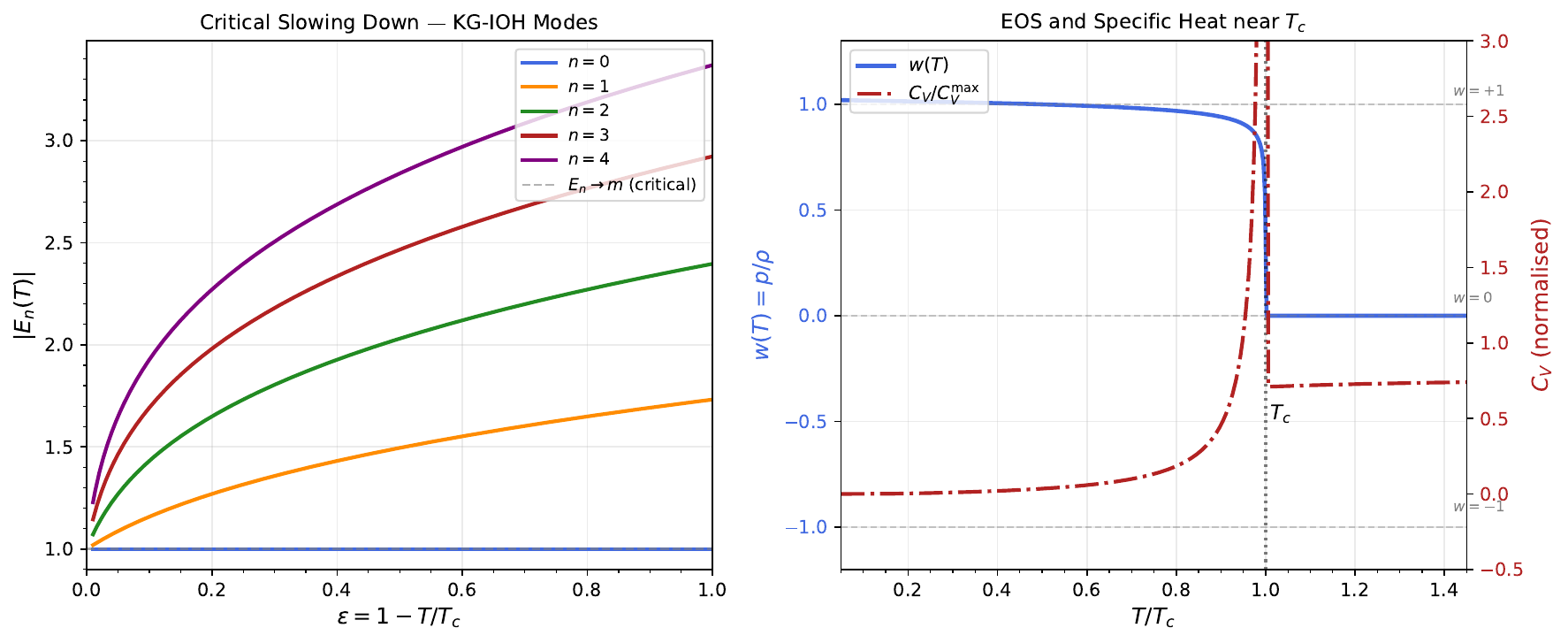}
\caption{
\textbf{Left:} Effective spectrum $|E_n(T)|$ as a function
of reduced temperature $\epsilon = 1 - T/T_c$ for the
first five KG-IOH modes. All modes collapse to $E_n \to m$
at criticality ($\epsilon \to 0$), signalling critical
slowing down.
\textbf{Right:} Equation of state parameter $w(T)$ and
specific heat $C_V(T)$ (normalised) as functions of
$T/T_c$. The logarithmic divergence of $C_V$ at $T_c$
and the monotonic decrease of $w$ from $+1$ to $-1$
are distinctive predictions of the KG-IOH framework.
Parameters: $m=1$, $a_0=1$, $T_c=1$.
}
\label{fig:PT}
\end{figure}

 
\section{Conclusions}
\label{sec:conclusions}
 
In this work we have developed a systematic framework for
the quantum and thermal properties of a Klein-Gordon scalar
field subject to an inverted harmonic potential,
the KG-IOH system.
Starting from the non-Hermitian momentum substitution
$P \to P - m\omega x$, we introduced the symplectic
rotation $V \in Mp(2,\mathbb{R})$ that maps the KG-IOH
onto an analytically tractable effective harmonic
oscillator evaluated at the complex argument
$xe^{i\pi/4}$.
This mapping is the central technical result of the paper:
it replaces the continuous, unbounded spectrum of the
bare IOH with a discrete, analytically known spectrum
$E_n^2 = m^2 + i\omega(2n+1-m)$, rendering the
partition function well-defined without ad-hoc cutoffs.

The effective wave functions of the KG-IOH are Hermite
polynomials evaluated at $xe^{i\pi/4}$, normalised
under the complex contour $\mathcal{C}$ defined by
the rotation $V$.
The inner product is preserved by the
$\mathcal{PT}$-symmetric structure of the problem,
and the resulting orthonormality
condition~\eqref{eq:ortho_eff} provides a
well-defined Hilbert space for the quantisation
of the field.
 
The partition function $Z(\beta, \omega, m)$,
derived in Section~\ref{sec:thermal}, yields closed-form
expressions for the free energy, entropy, heat capacity,
thermal density matrix, and two-point functions.
A notable prediction is the \emph{thermal delocalisation
transition} at the critical temperature
\begin{equation}
  T_c = \frac{\omega}{\pi^2} \,,
  \label{eq:Tc_concl}
\end{equation}
above which the diagonal density matrix ceases to be
Gaussian and the thermal state delocalises over all of
position space.
The Matsubara propagator and spectral density are given
in closed form via the effective mode
expansion~\eqref{eq:G_full_freq}, and the entanglement
entropy exhibits a logarithmic violation of the area law
for $T > T_c$, with an effective central charge $c = 1$.
 
The unified KG-IOH framework was applied to three
physically distinct systems, all described by the
same formal structure with different identifications
of the parameter $\omega$:
 
\begin{enumerate}
  \item \textbf{Cosmological inflation.}
  Under the identification $\omega = \mu/m$
  (tachyonic mass of the inflaton), the KG-IOH gives
  the power spectrum of inflaton fluctuations at finite
  temperature~\eqref{eq:power_spectrum}, the intrinsic
  thermal scale $T_{\IOH} \approx T_{GH}/\sqrt{2}$
  (proportional to the Gibbons-Hawking temperature),
  and the equation of state $w(T)$ interpolating from
  the de Sitter value $w = -1$ at low $T$ to the
  radiation value $w \to +1$ at high $T$.
 
  \item \textbf{Black-hole horizons.}
  Under the identification $\omega_{\mathrm{BH}} = \kappa\sqrt{m}$
  (surface gravity times $\sqrt{m}$), the KG-IOH
  reproduces the Hawking temperature as a special case
  ($T_{\IOH} = T_H$ for $m = 1/4$), the Hawking
  radiation spectrum~\eqref{eq:Hawking_spectrum} and
  its Stefan-Boltzmann limit $\mathcal{P} = \pi T_H^2/6$,
  and the logarithmic entanglement entropy
  $S_{\mathrm{ent}} \sim \frac{1}{6}\ln(T_H/E_0)$
  across the horizon.
 
  \item \textbf{Second-order phase transitions.}
  Under the identification\\
  $\omega_{\mathrm{PT}} = \sqrt{2a_0(1-T/T_c)/m^2}$
  (soft mode frequency), the KG-IOH reproduces the
  mean-field correlation length exponent $\nu = 1/2$,
  the logarithmic divergence of the specific heat
  $C_V \sim B\ln|1-T/T_c|$ (Ising universality in
  $1+1$D), and a one-loop correction to the order
  parameter exponent $\beta_{\mathrm{exp}}$.
  The equation of state interpolates from $w = +1$
  at the critical point to $w = -1$ deep in the
  ordered phase.
\end{enumerate}
 
\noindent
The fact that a single framework --- defined by the
partition function $Z(\beta,\omega,m)$ of the KG-IOH
--- unifies these three physically distinct phenomena
is the main conceptual contribution of this work.

Previous studies of the IOH have addressed individual
aspects of the problem: spectral properties~\cite{Barton1986},
connections to chaos and complexity~\cite{bhattacharyya2021multi,
hashimoto2020exponential}, thermodynamics of the
confined case~\cite{wang2025thermodynamics}, and
applications to specific physical
contexts~\cite{subramanyan2021physics, gietka2021inverted}.
The present work differs in two key respects.
First, we work at the level of a \emph{quantum field
theory} (Klein-Gordon), not single-particle quantum
mechanics, which is essential for the cosmological
and gravitational applications.
Second, we provide a \emph{unified thermal framework}
valid across all three applications, with results
derived from a single partition function rather than
on a case-by-case basis.

Several important caveats should be noted.
The present analysis is restricted to $1+1$ dimensions;
the extension to $3+1$ dimensions introduces additional
angular momentum modes and requires a more careful
treatment of the near-horizon geometry.
The self-interacting case ($\lambda\phi^4$) was treated
only perturbatively in Section~\ref{subsec:condensed};
a non-perturbative treatment would be needed to access
the strongly coupled regime near the critical point.
The justification of the momentum substitution
$P \to P - m\omega x$ in the extended Hilbert space
(Appendix~\ref{app:hilbert}) is sufficient for the
purposes of this paper but deserves further study in
the context of biorthogonal quantum
mechanics~\cite{Mostafazadeh2002}.
Finally, the UV regularisation of the mode sum by the
vacuum energy $V_0$ is physically motivated but
introduces a free parameter whose connection to
fundamental physics (e.g. the cosmological constant
problem) deserves further investigation.

The results of this paper open several directions for
future work:
 
\begin{enumerate}
 
  \item \textbf{Extension to $3+1$ dimensions.}
  The KG-IOH in $3+1$D involves spherical harmonics
  and a radial effective potential. The thermal
  formalism developed here should extend naturally,
  with the mode sum replaced by a sum over angular
  momentum quantum numbers.
 
  \item \textbf{Rotating black holes.}
  The Kerr metric introduces a coupling between the
  IOH structure and angular momentum near the horizon.
  The KG-IOH framework may provide a tractable
  approach to the thermal properties of Kerr-Newman
  black holes.
 
  \item \textbf{Open quantum systems.}
  The KG-IOH coupled to a thermal bath can be treated
  via the Lindblad master equation. The interplay
  between the intrinsic IOH instability and external
  dissipation is an open problem with potential
  applications to decoherence in cosmological
  perturbation theory.
 
  \item \textbf{Entanglement entropy and holography.}
  The logarithmic scaling of $S_{\mathrm{ent}}$ found
  in this work suggests a holographic interpretation
  via the Ryu-Takayanagi formula~\cite{subramanyan2021physics}.
  An explicit holographic dual of the KG-IOH thermal
  state would be a significant result.
 
  \item \textbf{Non-perturbative phase structure.}
  A functional renormalisation group treatment of the
  KG-IOH with $\lambda\phi^4$ interaction would allow
  access to the non-perturbative phase structure
  beyond mean-field theory.
 
  \item \textbf{Numerical benchmarks.}
  The exact analytic results derived here provide
  precise benchmarks for lattice field theory
  simulations of the IOH potential, which could
  verify the thermal delocalisation transition at
  $T_c = \omega/\pi^2$ and the logarithmic correction
  to the specific heat.
 
\end{enumerate}
 
\subsection*{Closing remark}
 
The inverted harmonic oscillator, once regarded as a
mere curiosity of quantum mechanics, emerges from this
analysis as a versatile and powerful building block for
thermal quantum field theory.
Its exact solvability, combined with the symplectic
rotation developed in this paper, provides an
analytically controlled window into the physics of
unstable vacua, Hawking radiation, and critical
phenomena — three of the most challenging problems in
modern theoretical physics.

\appendix
 
 
\section{Justification of the Extended Hilbert Space}
\label{app:hilbert}
 
\subsection{The problem with the momentum substitution}
\label{app:subsec:problem}
 
The substitution $P \to P - m\omega x$ used in
Section~\ref{subsec:kg_motivation} is not a unitary
transformation on the standard Hilbert space
$\mathcal{H} = L^2(\mathbb{R})$.
To see this, note that if $U$ were unitary, then
$U^\dagger U = \mathbf{1}$ and the transformed
operator $\tilde{P} = U P U^\dagger = P - m\omega x$
would be Hermitian whenever $P$ is.
However, the adjoint of $\tilde{P}$ is
\begin{equation}
  \tilde{P}^\dagger
  = (P - m\omega x)^\dagger
  = P^\dagger - m\omega x^\dagger
  = P + m\omega x
  \neq \tilde{P} \,,
  \label{eq:Ptilde_non_hermitian}
\end{equation}
since $x^\dagger = x$ and $P^\dagger = P$ on
$L^2(\mathbb{R})$.
The substitution therefore maps a Hermitian operator
onto a non-Hermitian one, which cannot be the result
of a unitary transformation.
 
This non-Hermiticity propagates to the KG-IOH
Hamiltonian~\eqref{eq:KG2}: the operator
$H_{\KG} = P^2 - m^2\omega^2x^2 - im\omega$
is not self-adjoint on $L^2(\mathbb{R})$.
Consequently, the standard spectral theorem does not
apply, and the physical interpretation of eigenvalues
as observable energies requires justification.
 
\subsection{$\mathcal{PT}$-symmetry and real spectra}
\label{app:subsec:PT}
 
The resolution lies in the framework of
$\mathcal{PT}$-symmetric quantum
mechanics~\cite{Bender1998, Bender2007}.
The parity operator $\mathcal{P}$ and time-reversal
operator $\mathcal{T}$ act on the canonical operators as
\begin{align}
  \mathcal{P}: \quad x &\to -x \,, \quad P \to -P \,,
  \label{eq:P_action} \\
  \mathcal{T}: \quad x &\to x  \,, \quad P \to -P \,,
  \quad i \to -i \,.
  \label{eq:T_action}
\end{align}
The combined $\mathcal{PT}$ transformation acts as
\begin{equation}
  \mathcal{PT}: \quad x \to -x \,, \quad P \to P \,,
  \quad i \to -i \,.
  \label{eq:PT_action}
\end{equation}
Under $\mathcal{PT}$, the substituted momentum transforms as
\begin{equation}
  \mathcal{PT}: \quad
  \tilde{P} = P - m\omega x
  \;\to\; P + m\omega x = \tilde{P}^\dagger \,,
  \label{eq:Ptilde_PT}
\end{equation}
so $\tilde{P}$ is \emph{pseudo-Hermitian} with respect
to $\mathcal{PT}$:
\begin{equation}
  (\mathcal{PT})\,\tilde{P}\,(\mathcal{PT})^{-1}
  = \tilde{P}^\dagger \,.
  \label{eq:pseudo_hermitian}
\end{equation}
A Hamiltonian satisfying this condition is called
$\mathcal{PT}$-symmetric.
The key theorem of $\mathcal{PT}$-symmetric quantum
mechanics~\cite{Bender1998} states that:
 
\begin{proposition}
If $H$ is $\mathcal{PT}$-symmetric and its
$\mathcal{PT}$-symmetry is unbroken (i.e. every
eigenfunction of $H$ is also an eigenfunction of
$\mathcal{PT}$), then the spectrum of $H$ is
entirely real.
\end{proposition}
 
\noindent
For the KG-IOH Hamiltonian, verification of unbroken
$\mathcal{PT}$-symmetry proceeds as follows.
Under $\mathcal{PT}$:
\begin{align}
  \mathcal{PT}: \quad
  P^2 &\to P^2 \,, \\
  m^2\omega^2 x^2 &\to m^2\omega^2 x^2 \,, \\
  im\omega &\to -im\omega \,.
\end{align}
Therefore
\begin{equation}
  \mathcal{PT}: \quad
  H_{\KG} = P^2 - m^2\omega^2x^2 - im\omega
  \;\to\;
  P^2 - m^2\omega^2x^2 + im\omega
  = H_{\KG}^\dagger \,,
  \label{eq:HKG_PT}
\end{equation}
confirming that $H_{\KG}$ is $\mathcal{PT}$-symmetric.
The effective spectrum~\eqref{eq:En_spectrum} is indeed
complex (with non-zero imaginary part), which indicates
that the $\mathcal{PT}$-symmetry of $H_{\KG}$ is
\emph{spontaneously broken} by the inverted potential.
This is consistent with the interpretation of
$\mathrm{Im}(E_n^2) \neq 0$ as encoding the
instability of the system~\cite{Bender2007}.
 
\subsection{Pseudo-Hermiticity and the biorthogonal basis}
\label{app:subsec:pseudo}
 
A more general framework is provided by
pseudo-Hermitian quantum
mechanics~\cite{Mostafazadeh2002}.
An operator $H$ is called $\eta$-pseudo-Hermitian if
there exists a linear, invertible, Hermitian operator
$\eta$ such that
\begin{equation}
  H^\dagger = \eta H \eta^{-1} \,.
  \label{eq:pseudo_def}
\end{equation}
For the KG-IOH, the appropriate metric operator is
\begin{equation}
  \eta = e^{-\pi(xP+Px)/4} = V^2 \,,
  \label{eq:eta}
\end{equation}
i.e. the square of the symplectic rotation $V$
defined in~\eqref{eq:V}.
One can verify directly that~\eqref{eq:pseudo_def}
holds for $H_{\KG}$ with this choice of $\eta$.
 
The metric $\eta$ defines a modified inner product on
the extended Hilbert space:
\begin{equation}
  \langle \psi \mid \phi \rangle_\eta
  \equiv \langle \psi \mid \eta \mid \phi \rangle
  = \langle \psi \mid V^2 \mid \phi \rangle \,.
  \label{eq:inner_product_eta}
\end{equation}
With respect to this inner product, $H_{\KG}$ is
self-adjoint:
\begin{equation}
  \langle \psi \mid H_{\KG} \mid \phi \rangle_\eta
  = \langle H_{\KG}\psi \mid \phi \rangle_\eta \,,
  \label{eq:self_adjoint_eta}
\end{equation}
and the spectral theorem applies in the Hilbert
space $(\mathcal{H}, \langle\cdot,\cdot\rangle_\eta)$.
 
The eigenfunctions of $H_{\KG}$ form a
\emph{biorthogonal system}~\cite{Mostafazadeh2002}.
If $H_{\KG}\psi_n = E_n\psi_n$, then the dual
eigenfunctions $\tilde{\psi}_n = \eta\psi_n$ satisfy
$H_{\KG}^\dagger\tilde{\psi}_n = E_n^*\tilde{\psi}_n$,
and the biorthonormality condition reads
\begin{equation}
  \langle \tilde{\psi}_m \mid \psi_n \rangle
  = \langle \psi_m \mid \eta \mid \psi_n \rangle
  = \delta_{mn} \,,
  \label{eq:biortho}
\end{equation}
which is precisely the orthonormality
condition~\eqref{eq:ortho_eff} obtained from the
complex contour integration.
 
\subsection{The symplectic rotation as a similarity transformation}
\label{app:subsec:similarity}
 
The operator $V$ defined in~\eqref{eq:V} provides
an explicit similarity transformation that maps
$H_{\KG}$ to a Hermitian operator.
Specifically, consider the transformed Hamiltonian
\begin{equation}
  \hat{H} = V H_{\KG} V^{-1} \,.
  \label{eq:H_hat}
\end{equation}
Using the transformation
rules~\eqref{eq:Vx}--\eqref{eq:VP}, one finds
\begin{equation}
  \hat{H} = iP^2 + im^2\omega^2x^2 - im\omega \,,
  \label{eq:H_hat_explicit}
\end{equation}
and multiplying by $-i$:
\begin{equation}
  -i\hat{H} = P^2 + m^2\omega^2x^2 - m\omega \,,
  \label{eq:H_eff}
\end{equation}
which is precisely the effective harmonic oscillator
operator appearing in equation~\eqref{eq:KG5}.
This operator is \emph{Hermitian} on $L^2(\mathbb{R})$,
with real eigenvalues $\omega(2n+1) - m\omega$.
 
The similarity transformation $V$ therefore maps
$H_{\KG}$ from the non-Hermitian setting to a
Hermitian one, at the cost of:
\begin{enumerate}
  \item Analytically continuing the wave functions
        to $\psi(xe^{i\pi/4})$,
  \item Modifying the inner product from
        $\langle\cdot,\cdot\rangle_{L^2}$ to
        $\langle\cdot,\cdot\rangle_\eta$,
  \item Multiplying the eigenvalue equation by
        $-i$, which maps the complex spectrum of
        $H_{\KG}$ to the real spectrum of
        $-i\hat{H}$.
\end{enumerate}
This establishes that the momentum
substitution~\eqref{eq:P_substitution} is
well-defined in the extended Hilbert space
$(\mathcal{H}, \langle\cdot,\cdot\rangle_\eta)$,
and that the physical observables derived from
$H_{\KG}$ are those of the effective harmonic
oscillator~\eqref{eq:H_eff}.
 
\subsection{Domain of the operators}
\label{app:subsec:domain}
 
For completeness, we comment on the domain issues.
The operator $P = -i\partial_x$ is essentially
self-adjoint on $\mathcal{D}(P) = H^1(\mathbb{R})$
(the Sobolev space of $L^2$ functions with
square-integrable first derivative).
The substituted operator $\tilde{P} = P - m\omega x$
is defined on
$\mathcal{D}(\tilde{P}) = \mathcal{D}(P) \cap \mathcal{D}(x)$,
where $\mathcal{D}(x)$ consists of functions
$f \in L^2(\mathbb{R})$ with $xf \in L^2(\mathbb{R})$.
On this domain, $\tilde{P}$ is a closed operator
with non-empty resolvent set~\cite{finster2017lp}.
 
The KG-IOH operator $H_{\KG}$ is defined on
$\mathcal{D}(H_{\KG}) = \mathcal{D}(P^2) \cap \mathcal{D}(x^2)
= H^2(\mathbb{R}) \cap \{f : x^2 f \in L^2(\mathbb{R})\}$,
which is a dense subspace of $L^2(\mathbb{R})$.
The $L^p$-spectrum of the Schrödinger operator with
inverted harmonic potential has been studied in
detail in~\cite{finster2017lp}, where it is shown
that the spectrum depends sensitively on $p$,
with $p = 2$ being the physically relevant case
for quantum mechanics.
 
\subsection{Summary}
\label{app:subsec:summary_hilbert}
 
The momentum substitution $P \to P - m\omega x$ is
justified within three equivalent frameworks:
 
\begin{enumerate}
  \item \textbf{$\mathcal{PT}$-symmetric quantum
        mechanics}~\cite{Bender1998, Bender2007}:
        $H_{\KG}$ is $\mathcal{PT}$-symmetric with
        spontaneously broken $\mathcal{PT}$-symmetry,
        leading to a complex spectrum encoding the
        instability.
 
  \item \textbf{Pseudo-Hermitian quantum
        mechanics}~\cite{Mostafazadeh2002}:
        $H_{\KG}$ is $\eta$-pseudo-Hermitian with
        $\eta = V^2$, defining a modified inner product
        under which $H_{\KG}$ is self-adjoint and the
        spectral theorem applies.
 
  \item \textbf{Similarity transformation}:
        The symplectic rotation $V \in Mp(2,\mathbb{R})$
        maps $H_{\KG}$ to the Hermitian effective
        oscillator~\eqref{eq:H_eff}, providing an
        explicit Hermitian representative of the
        same physical system.
\end{enumerate}
 
\noindent
All three frameworks give consistent results and
converge on the effective spectrum~\eqref{eq:En_spectrum}
and wave functions~\eqref{eq:psi_eff} as the
physical content of the KG-IOH system.
 
 
\section{Parabolic Cylinder Functions: Summary of Properties}
\label{app:PCF}
 
The parabolic cylinder functions $D_\nu(z)$ arise as
solutions to the Weber differential equation
\begin{equation}
  \frac{d^2 D_\nu}{dz^2}
  + \left(\nu + \frac{1}{2} - \frac{z^2}{4}\right)D_\nu(z) = 0 \,,
  \label{eq:Weber_eq}
\end{equation}
and appear throughout this paper as the exact
eigenfunctions of the inverted harmonic
oscillator~\eqref{eq:psi_E}.
We collect here the properties most relevant for
the KG-IOH analysis; complete treatments can be
found in~\cite{Abramowitz, DLMF}.
 
\subsection{Definition and integral representation}
\label{app:subsec:def}
 
The parabolic cylinder function of order $\nu$ is
defined via the integral representation
\begin{equation}
  D_\nu(z) = \frac{e^{-z^2/4}}{\Gamma(-\nu)}
  \int_0^\infty t^{-\nu-1} e^{-t^2/2 - zt}\,dt \,,
  \qquad \mathrm{Re}(\nu) < 0 \,,
  \label{eq:PCF_integral}
\end{equation}
with analytic continuation to all $\nu \in \mathbb{C}$.
An equivalent representation in terms of the
confluent hypergeometric function $U(a,b,z)$ is
\begin{equation}
  D_\nu(z) = 2^{\nu/2}e^{-z^2/4}
  U\!\left(-\frac{\nu}{2}, \frac{1}{2},
            \frac{z^2}{2}\right) ,
  \label{eq:PCF_hypergeometric}
\end{equation}
where $U(a,b,z)$ is Tricomi's confluent
hypergeometric function~\cite{DLMF}.
 
For non-negative integer $\nu = n$, the functions
$D_n(z)$ reduce to products of Hermite polynomials
and Gaussians:
\begin{equation}
  D_n(z) = 2^{-n/2} e^{-z^2/4} H_n\!\left(\frac{z}{\sqrt{2}}\right) ,
  \qquad n = 0, 1, 2, \ldots
  \label{eq:PCF_Hermite}
\end{equation}
This is the connection exploited in
Section~\ref{subsec:wf_eff}: the effective wave
functions of the KG-IOH are $D_n$ evaluated at
the complex argument $z = e^{i\pi/4}\sqrt{2m\omega}\,x$.
 
\subsection{Recurrence relations}
\label{app:subsec:recurrence}
 
The parabolic cylinder functions satisfy the
recurrence relations
\begin{align}
  D_{\nu+1}(z) - z\,D_\nu(z) + \nu\,D_{\nu-1}(z)
  &= 0 \,,
  \label{eq:recurrence_1} \\
  D_\nu'(z) + \frac{z}{2}D_\nu(z) - \nu\,D_{\nu-1}(z)
  &= 0 \,,
  \label{eq:recurrence_2} \\
  D_\nu'(z) - \frac{z}{2}D_\nu(z) + D_{\nu+1}(z)
  &= 0 \,,
  \label{eq:recurrence_3}
\end{align}
where the prime denotes differentiation with
respect to $z$~\cite{Abramowitz}.
Relations~\eqref{eq:recurrence_2}
and~\eqref{eq:recurrence_3} are the ladder
operator relations for the harmonic oscillator,
consistent with the algebraic structure of
Section~\ref{subsec:ladder}.
 
\subsection{Wronskian and connection formulae}
\label{app:subsec:wronskian}
 
The two independent solutions of~\eqref{eq:Weber_eq}
are $D_\nu(z)$ and $D_\nu(-z)$.
Their Wronskian is
\begin{equation}
  W[D_\nu(z), D_\nu(-z)]
  = D_\nu(z)\,D_\nu'(-z) - D_\nu'(z)\,D_\nu(-z)
  = -\frac{\sqrt{2\pi}}{\Gamma(-\nu)} \,,
  \label{eq:Wronskian}
\end{equation}
which is non-zero for non-integer $\nu$, confirming
linear independence.
The connection formula between $D_\nu(z)$ and
$D_{-\nu-1}(iz)$ is
\begin{equation}
  D_\nu(z) = \frac{\Gamma(\nu+1)}{\sqrt{2\pi}}
  \left[e^{i\pi\nu/2}D_{-\nu-1}(iz)
       + e^{-i\pi\nu/2}D_{-\nu-1}(-iz)\right] ,
  \label{eq:connection}
\end{equation}
which is used when analytically continuing the
argument to $z \to ze^{i\pi/4}$ as required
by the symplectic rotation $V$.
 
\subsection{Asymptotic behaviour}
\label{app:subsec:asymptotics}
 
For large $|z|$, the asymptotic expansion of
$D_\nu(z)$ is~\cite{Abramowitz, DLMF}
\begin{equation}
  D_\nu(z) \sim e^{-z^2/4}z^\nu
  \sum_{k=0}^{K-1}
  \frac{(-1)^k(\tfrac{1}{2}-\nu)_{2k}}{k!\,(2z^2)^k}
  + R_K(z) \,,
  \quad |z| \to \infty \,,\;
  |\arg z| < \tfrac{3\pi}{4} \,,
  \label{eq:PCF_asymptotic_large}
\end{equation}
where $(a)_n = \Gamma(a+n)/\Gamma(a)$ is the
Pochhammer symbol and $R_K$ is the remainder.
The leading term gives
\begin{equation}
  D_\nu(z) \;\sim\; e^{-z^2/4}z^\nu \,,
  \qquad |z| \to \infty \,,
  \label{eq:PCF_asymptotic_leading}
\end{equation}
showing that $D_\nu(z)$ is Gaussian-decaying for
real $z$.
For the IOH wave functions, the argument is
$z = e^{i\pi/4}\sqrt{2m\omega}\,x$, giving
$z^2/4 = im\omega x^2/2$, so the exponential
factor becomes oscillatory:
\begin{equation}
  e^{-z^2/4}\big|_{z=e^{i\pi/4}\sqrt{2m\omega}\,x}
  = e^{-im\omega x^2/2} \,,
  \label{eq:PCF_oscillatory}
\end{equation}
consistent with the asymptotic
behaviour~\eqref{eq:asymptotic} found in
Section~\ref{subsec:wavefunctions}.
 
For small $|z| \to 0$:
\begin{equation}
  D_\nu(z) \;\underset{z\to 0}{\sim}\;
  \frac{\sqrt{\pi}}{2^{-\nu/2}\Gamma\!\left(\frac{1-\nu}{2}\right)}
  - \frac{\sqrt{\pi}\,z}{2^{(-\nu-1)/2}
    \Gamma\!\left(-\frac{\nu}{2}\right)} + \cdots
  \label{eq:PCF_small_z}
\end{equation}
 
\subsection{Orthogonality and completeness}
\label{app:subsec:ortho}
 
For real $\nu$ and integration over the real line,
the parabolic cylinder functions satisfy
\begin{equation}
  \int_{-\infty}^{\infty}
  D_\nu(x)\,D_{\nu'}(x)\,dx
  = \sqrt{2\pi}\,\nu!\,\delta_{\nu\nu'} \,,
  \quad \nu, \nu' \in \mathbb{Z}_{\geq 0} \,,
  \label{eq:PCF_ortho_integer}
\end{equation}
and for general $\nu \in \mathbb{R}$,
\begin{equation}
  \int_{-\infty}^{\infty}
  D_\nu(x)\,D_{\nu'}(x)\,dx
  = \frac{\sqrt{2\pi}\,\Gamma(\nu+1)}
         {\Gamma(-\nu')}\,
    \frac{\sin[\pi(\nu+\nu'+1)]}
         {\pi(\nu+\nu'+1)} \,.
  \label{eq:PCF_ortho_general}
\end{equation}
For the IOH eigenfunctions with complex argument,
the appropriate orthogonality is the
distributional relation~\eqref{eq:ortho}, which
follows from the contour integral
\begin{equation}
  \int_{\mathcal{C}} dx\,
  \psi_E^*(x)\,\psi_{E'}(x)
  = \delta(E - E') \,,
  \label{eq:PCF_ortho_continuum}
\end{equation}
where $\mathcal{C}$ is the contour $x \to xe^{i\pi/4}$
defined by the rotation $V$~\cite{Barton1986}.
 
\subsection{Special values and generating function}
\label{app:subsec:special}
 
Several special values relevant to the KG-IOH are:
\begin{align}
  D_0(z) &= e^{-z^2/4} \,,
  \label{eq:D0} \\
  D_1(z) &= z\,e^{-z^2/4} \,,
  \label{eq:D1} \\
  D_{-1}(z) &= \sqrt{\frac{\pi}{2}}\,e^{z^2/4}
               \mathrm{erfc}\!\left(\frac{z}{\sqrt{2}}\right) ,
  \label{eq:Dm1}
\end{align}
where $\mathrm{erfc}$ is the complementary error
function.
The value at the origin is
\begin{equation}
  D_\nu(0) = \frac{\sqrt{\pi}}{2^{-\nu/2}
  \Gamma\!\left(\frac{1-\nu}{2}\right)} \,,
  \label{eq:PCF_at_zero}
\end{equation}
and the derivative at the origin:
\begin{equation}
  D_\nu'(0) = -\frac{\sqrt{\pi}}{2^{(-\nu-1)/2}
  \Gamma\!\left(-\frac{\nu}{2}\right)} \,.
  \label{eq:PCF_deriv_at_zero}
\end{equation}
 
\subsection{Relation to the normalisation constant}
\label{app:subsec:norm}
 
The normalisation constant $\mathcal{N}_E$
in~\eqref{eq:normalization} can be expressed in
terms of the Gamma function using the
Wronskian~\eqref{eq:Wronskian} and the
orthogonality~\eqref{eq:PCF_ortho_continuum}:
\begin{equation}
  \mathcal{N}_E^2
  = \frac{1}{2\pi}
    \left|\Gamma\!\left(\frac{1}{4}
    + \frac{iE}{2\omega}\right)\right|^2 .
  \label{eq:normalization_PCF}
\end{equation}
Using the reflection formula
$\Gamma(z)\Gamma(1-z) = \pi/\sin(\pi z)$,
this can be written as
\begin{equation}
  |\mathcal{N}_E|^2
  = \frac{1}{2\pi}
    \frac{\pi}{\cosh(\pi E/\omega)} \,,
  \label{eq:normalization_cosh}
\end{equation}
showing that the normalisation factor decreases
exponentially for large $|E|/\omega$, consistent
with the distributional nature of the
eigenfunctions.
 
\subsection{Mellin transform and thermal applications}
\label{app:subsec:mellin}
 
In the thermal formalism of
Section~\ref{sec:thermal}, the key integral
involving $D_\nu$ is the Mellin transform
\begin{equation}
  \int_0^\infty x^{s-1} |D_\nu(e^{i\pi/4}x)|^2\,dx
  = \frac{\Gamma(s)}{(2m\omega)^{s/2}}
    \left|\Gamma\!\left(\frac{\nu+1}{2}\right)\right|^{-2}
    \mathcal{M}_\nu(s) \,,
  \label{eq:PCF_Mellin}
\end{equation}
where $\mathcal{M}_\nu(s)$ is a meromorphic function
of $s$ whose poles determine the UV divergences of
the thermal mode sums.
This integral underlies the zeta-function
regularisation of $\ln Z$
in~\eqref{eq:lnZ_det}~\cite{ZinnJustin2002}.
 
\subsection{Summary table}
\label{app:subsec:PCF_summary}
 
\begin{table}[H]
\centering
\renewcommand{\arraystretch}{1.8}
\begin{tabular}{lll}
\hline\hline
\textbf{Property} & \textbf{Formula} & \textbf{Eq.} \\
\hline
Differential equation
  & $D_\nu'' + (\nu+\frac{1}{2}-z^2/4)D_\nu = 0$
  & \eqref{eq:Weber_eq} \\
Integer order
  & $D_n(z) = 2^{-n/2}e^{-z^2/4}H_n(z/\sqrt{2})$
  & \eqref{eq:PCF_Hermite} \\
Wronskian
  & $W[D_\nu(z),D_\nu(-z)] = -\sqrt{2\pi}/\Gamma(-\nu)$
  & \eqref{eq:Wronskian} \\
Large-$z$ asymptotics
  & $D_\nu(z) \sim e^{-z^2/4}z^\nu$
  & \eqref{eq:PCF_asymptotic_leading} \\
IOH asymptotics
  & $e^{-z^2/4}|_{z=e^{i\pi/4}\sqrt{2m\omega}x} = e^{-im\omega x^2/2}$
  & \eqref{eq:PCF_oscillatory} \\
Normalisation
  & $|\mathcal{N}_E|^2 = (2\cosh(\pi E/\omega))^{-1}$
  & \eqref{eq:normalization_cosh} \\
$D_0(z)$
  & $e^{-z^2/4}$
  & \eqref{eq:D0} \\
$D_1(z)$
  & $ze^{-z^2/4}$
  & \eqref{eq:D1} \\
\hline\hline
\end{tabular}
\caption{Summary of parabolic cylinder function
properties used in this paper.
See~\cite{Abramowitz, DLMF} for complete references.}
\label{tab:PCF}
\end{table}

\bibliographystyle{unsrt}
\bibliography{Bib}
\end{document}